\documentclass[structabstract]{aa}
\usepackage{float}
\usepackage{graphicx}
\usepackage{natbib}
\usepackage{tabularx}
\usepackage{txfonts}
\usepackage{hyperref}
\hypersetup{
    colorlinks,
    citecolor=blue,
    filecolor=blue,
    linkcolor=blue,
    urlcolor=blue,
    menucolor=black 
}
\bibpunct{(}{)}{;}{a}{}{,}
\begin{document}

\newcommand{\ir}{\hat{I}_{r}}
\newcommand{\ia}{\hat{I}_{\phi}}
\newcommand{\irm}{\left\langle\ir\right\rangle}
\newcommand{\iam}{\left\langle\ia\right\rangle}
\newcommand{\sir}{\int\ir{}dr}
\newcommand{\sia}{{\rm RMS}_{\ia-\iam}}
\newcommand{\psfone}{K_1}
\newcommand{\psftwo}{K_2}
\newcommand{\inorm}{I/I_0}
\newcommand{\vl}{\left\langle{}\bf{v}_{\rm l}\right\rangle}
\newcommand{\dvl}{\Delta\vl_{\rm QS}}
\newcommand{\vbl}{\left\langle{}B_{\rm l}\right\rangle}
\newcommand{\bl}{|\left\langle{}B_{\rm l}\right\rangle|}
\newcommand{\bmu}{\bl/\mu}
\newcommand{\mbl}{\epsilon_{\bl}}
\newcommand{\noise}{\sigma_{\vbl}}
\newcommand{\mblmag}{\epsilon_{\bl{\rm ,mag}}}
\newcommand{\dbl}{\Delta\mbl}
\newcommand{\nmag}{n_{{\rm mag}}}
\newcommand{\dflux}{\Delta\left(\nmag\mblmag\right)}
\newcommand{\iqs}{\left\langle \inorm\right\rangle_{\rm QS}}
\newcommand{\iqp}{\left(\inorm\right)_{\rm QS,P}}
\newcommand{\ipu}{\left(\inorm\right)_{\rm P,U}}
\newcommand{\vobs}{\bf{v}_{\rm SDO}}
\newcommand{\vrot}{\bf{v}_{\rm rot}}
\newcommand{\mps}{{\rm ms^{-1}}}

\title{Point spread function of SDO/HMI and the effects of stray light correction on the apparent properties of solar surface phenomena}
\author{K.~L.~Yeo\inst{\ref{inst1},\ref{inst2}}\and A.~Feller\inst{\ref{inst1}}\and S.~K.~Solanki\inst{\ref{inst1},\ref{inst3}}\and S.~Couvidat\inst{\ref{inst4}}\and S.~Danilovic\inst{\ref{inst1}}\and N.~A.~Krivova\inst{\ref{inst1}}}
\institute{
Max-Planck-Institut f\"{u}r Sonnensystemforschung, Max-Planck-Stra\ss{}e 2, 37191 Katlenburg-Lindau, Germany \\
\email{yeo@mps.mpg.de}
\label{inst1}
\and
Technische Universit\"a{}t Braunschweig, Institut f\"{u}r Geophysik und Extraterrestrische Physik, Mendelssohnstra\ss{}e 3, 38106 Braunschweig, Germany
\label{inst2}
\and
School of Space Research, Kyung Hee University, Yongin, 446-701 Gyeonggi, Korea
\label{inst3}
\and
W.W. Hansen Experimental Physics Laboratory, Stanford University, Stanford, CA 94305-4085, USA
\label{inst4}
}
\date{Received 19 August 2013 / Accepted 15 October 2013}
\abstract{
}{We present a point spread function (PSF) for the Helioseismic and Magnetic Imager (HMI) onboard the Solar Dynamics
Observatory (SDO) and discuss the effects of its removal on the apparent properties of solar surface phenomena in HMI data.
}{The PSF was retrieved from observations of Venus in transit by matching it to the convolution of a model of the venusian disc and solar background with a guess PSF. We described the PSF as the sum of five Gaussian functions, the amplitudes of which vary sinusoidally with azimuth. This relatively complex functional form was required by the data. Observations recorded near in time to the transit of Venus were corrected for instrumental scattered light by the deconvolution with the PSF. We also examined the variation in the shape of the solar aureole in daily data, as an indication of PSF changes over time.
}{Granulation contrast in restored HMI data is greatly enhanced relative to the original data and exhibit reasonable agreement with numerical simulations. Image restoration enhanced the apparent intensity and pixel averaged magnetic field strength of photospheric magnetic features significantly. For small-scale magnetic features, restoration enhanced intensity contrast in the continuum and core of the Fe I 6173 \AA{} line by a factor of 1.3, and the magnetogram signal by a factor of 1.7. For sunspots and pores, the enhancement varied strongly within and between features, being more acute for smaller features. Magnetic features are also rendered smaller, as signal smeared onto the surrounding quiet Sun is recovered. Image restoration increased the apparent amount of magnetic flux above the noise floor by a factor of about 1.2, most of the gain coming from the quiet Sun. Line-of-sight velocity due to granulation and supergranulation is enhanced by a factor of 1.4 to 2.1, depending on position on the solar disc. The shape of the solar aureole varied, with time and between the two CCDs. There are also indications that the PSF varies across the FOV. However, all these variations were found to be relatively small, such that a single PSF can be applied to HMI data from both CCDs, over the period examined without introducing significant error.
}{Restoring HMI observations with the PSF presented here returns a reasonable estimate of the stray light-free intensity contrast. Image restoration affects the measured radiant, magnetic and dynamic properties of solar surface phenomena sufficiently to significantly impact interpretation.
}
\titlerunning{PSF and stray light correction of SDO/HMI data}
\authorrunning{K. L. Yeo et al.}
\maketitle

\section{Introduction}
\label{introduction}

Solar telescopes, like any real optical system, diverge from diffraction-limited behaviour due to optical aberrations and, in the case of ground-based instruments, the influence of the Earth's atmosphere. Optical aberrations arise from factors such as design and material constraints, imperfections in the fabrication, presence of impurities, thermal changes and jitter, and are practically unavoidable. Due to aperture diffraction and optical aberrations, radiation entering a given optical system is not entirely confined to the intended area on the focal plane but instead spread out, as described mathematically by the point spread function, PSF. This image blurring or loss of contrast is the so-termed stray light.

The apparent properties of solar phenomena is sensitive to stray light, accounting for its influence on solar imagery is necessary for proper interpretation and comparison with numerical models. This has been demonstrated repeatedly in the literature, for example, in studies looking at the limb darkening function \citep{pierce77,neckel94}, the intensity contrast of granulation \citep{sanchezcuberes00,danilovic08,wedemeyerbohm09,afram11}, and the brightness of small-scale magnetic concentrations \citep{title96,schnerr11} and sunspots \citep{albregtsen81,mathew07}.

Sophisticated techniques to correct solar observations for instrumental and atmospheric effects exist, the most common being speckle interferometry \citep{deboer92,vonderluhe93} and phase diversity \citep{gonsalves82,lofdahl94} methods. For spaceborne instruments, where variable atmospheric seeing is not a factor, a more conventional approach is often sufficient. Specifically, inferring the PSF from the distribution of intensity about the boundary between the bright and dark parts of partially illuminated scenes (such as, of the solar limb, and of transits of the Moon, Venus or Mercury) and restoring data by the deconvolution with it. Recent examples include the work of \cite{mathew07} with observations from SoHO/MDI\footnote{The Michelson Doppler Imager onboard the Solar and Heliospheric Observatory \citep{scherrer95}.}, \cite{wedemeyerbohm08}, \cite{wedemeyerbohm09} and \cite{mathew09} with Hinode/SOT/BFI\footnote{The Broadband Filter Imager of the Solar Optical Telescope onboard Hinode \citep{kosugi07}.}, and \cite{deforest09} with TRACE\footnote{The Transition Region And Coronal Explorer \citep{handy98}.} . 

In this paper we present an estimate of the PSF of the Helioseismic and Magnetic Imager onboard the Solar Dynamics Observatory, SDO/HMI \citep{schou12}. The PSF was derived from observations of Venus in transit. We also demonstrate the effects of correcting HMI data for stray light with this PSF on the apparent properties of various photospheric phenomena.

This study broadly follows the approach taken with the other spaceborne instruments listed above. It departs from these earlier efforts in that we constrain the PSF not only in the radial dimension but also in the azimuthal direction, recovering the anisotropy. This we will show to be crucial for accurate stray light removal (Sect. \ref{imagerestoration}).

The relationship between the radiance of magnetic features in the photosphere, and their size and position on the solar disc, is an important consideration in understanding and modelling the variation in solar irradiance \citep{domingo09}. HMI returns continuous, seeing-free, full-disc observations of intensity, Doppler shift and magnetic field at a constant, intermediate spatial resolution ($\sim1\:\rm{arcsec}$) and at relatively low noise. This renders it a suitable tool for constraining the radiant behaviour of photospheric magnetic features \citep{yeo13}. It is therefore of interest to enhance the quality of HMI observations by quantifying the stray light performance of the instrument. This would be of utility not only for the accurate examination of the radiant behaviour of magnetic features in HMI data but also any application that can benefit from stray light-free measurements of intensity, line-of-sight velocity and magnetic flux density.

The PSF presented here is, to our knowledge, the first on-orbit measurement of the stray light of HMI reported in the literature \citep[see][for the pre-launch measurement]{wachter12}. This is necessary given that the exact operating conditions of the sensor cannot be exactly simulated on the ground. Also, the stray light of the HMI might have changed from the time of the pre-launch calibration from changes in the condition of the instrument.

The HMI comprises of two identical $4096\times4096$ pixel CCD cameras, denoted `side' and `front'. The PSF was retrieved from images recorded on the side CCD during the transit of Venus on June 5 to 6, 2012. In addition to this transit of Venus, the HMI has also witnessed several partial lunar eclipses (seven, as of the end of 2012). In Sect. \ref{psl} we discuss the reasons for choosing the observations of Venus in transit over data from the partial lunar eclipses or of the solar limb for constraining the PSF, even though Venus has an atmosphere which had to be taken into account in deriving the PSF, introducing additional complexity to the task and uncertainty to the final estimate of the PSF.

In the following section, we detail the data selection (Sect. \ref{dataselection}), the PSF derivation (Sect. \ref{psfderivation}) and image restoration method (Sect. \ref{imagerestoration}), and how we accounted for the influence of Venus' atmosphere (Sect. \ref{venusatm}). Then, we verify the utility of the PSF presented here for image restoration, comparing the apparent granulation contrast in restored HMI observations and synthetic intensity maps generated from numerical simulation (Sect. \ref{gc}). We illustrate the result of image restoration on the intensity, Dopplergram and magnetogram data products of the instrument, looking at its effect on the following.
\begin{itemize}
	\item The intensity and magnetic field strength of small-scale magnetic concentrations (Sect. \ref{ssmcss1}).
	\item The intensity and magnetic field strength of sunspots and pores (Sect. \ref{ssmcss2}).
	\item The amount of magnetic flux on the solar surface (Sect. \ref{ssmcss3}).
	\item Line-of-sight velocity (Sect. \ref{ssmcss4}).
\end{itemize}
The retrieved PSF represents the stray light behaviour of the side CCD at the time of the transit of Venus, at the position in the field-of-view (FOV) occupied by the Venus disc. In Sect. \ref{psfdep} we examine the applicability of this PSF to other positions in the FOV, and to observations from the front CCD as well as from other times. Finally, a summary of the study is given in Sect. \ref{summary}.

\section{PSF derivation}
\label{method}

\subsection{Data selection}
\label{dataselection}

The HMI is a full-Stokes capable filtergram instrument. The instrument records full-disc polarimetric filtergrams continuously, at 3.75-s cadence, on the two identical CCDs. The filtergram sequence of the side CCD alternates between six polarizations ($I\pm{}Q$, $I\pm{}U$ and $I\pm{}V$) and six positions across the Fe I 6173 \AA{} line (at $\pm34$, $\pm103$ and $\pm172\:{\rm m\AA{}}$ from line centre). A set of 36 filtergrams, of each polarization at each line position, is collected every 135-s. The front CCD collects a set of 12 filtergrams, covering the Stokes $I+V$ and $I-V$ polarizations at the same line positions, every 45-s.

Dopplergrams, longitudinal magnetograms and intensity (continuum, and line depth and width) images, collectively termed the line-of-sight data products, are generated from the filtergram sequence of the side CCD at 720-s intervals, and from that of the front CCD at 45-s intervals. Stokes parameters (I, Q, U and V) and the corresponding Milne-Eddington inversion \citep{borrero11} are also produced, at 720-s cadence, from the filtergram sequence of the side CCD.

During the transit of Venus on June 5 to 6, 2012, the side CCD recorded filtergrams in the nearby continuum ($-344\:{\rm m\AA{}}$ from line centre), instead of the regular filtergram sequence. For the purpose of estimating the PSF of the instrument, we examined the $854\times854$ pixel crop, centred on the Venus disc, of 249 continuum filtergrams collected between second and third contact (i.e., the period the venusian disc was entirely within the solar disc). Care was taken to avoid filtergrams with pixels with spurious signal levels, a result of cosmic ray hits, on the venusian disc. The pixel scale was 0.504 arcsec/pixel. A summary description of this and the other HMI data employed in this study is given in Table \ref{hmidata}.

\begin{table*}
\caption{Summary description of the HMI data employed in this study and the sections in which their analysis is detailed.}
\label{hmidata}
\centering
\begin{tabularx}{\textwidth}{l>{\hsize=1.6\hsize}X>{\hsize=.7\hsize}X>{\hsize=.7\hsize}Xl}
\hline\hline
Index & Description & UTC time of observation & Denotation & Section(s)\\
\hline
1 & $854\times854$ pixel crop, centred on the venusian disc, of 249 continuum ($-344\:{\rm m\AA{}}$ from line centre) filtergrams recorded on the side CCD during the transit of Venus, between second and third contact. & Between 22:30, June 5 and 04:14, June 6, 2012. & & \ref{dataselection}\\
2 & The mean of 42 of the 249 images in item 1, between which the spatial distribution of intensity on the venusian disc is relatively similar. & Between 02:04 and 02:46, June 5, 2012. & Mean transit image & \ref{dataselection}, \ref{psfderivation}, \ref{venusatm}\\
3 & One of the 42 images used to produce item 2. & 02:25:37, June 6, 2012. & Test transit image & \ref{imagerestoration}, \ref{venusatm}\\
4 & $854\times854$ pixel crop, centred on the venusian disc, of a continuum side CCD filtergram recorded just before second contact. & 22:25:33, June 5, 2012. & Ingress image & \ref{venusatm}\\
5 & Continuum side CCD filtergram taken right after the venusian disc exited the solar disc completely. & 04:35:59, June 6, 2012. & Test continuum filtergram & \ref{gc}, \ref{psfdep1}\\
6 & 45-s longitudinal magnetogram from the front CCD closest in time ($<$ 1 minute) to item 5. & 04:35:22, June 6, 2012. & & \ref{gc}, \ref{psfdep1}\\
7 & A set of simultaneous 720-s Dopplergram, longitudinal magnetogram, line depth and continuum intensity images from the side CCD, recorded about an hour after the end of the transit of Venus, when said CCD resumed collection of the regular filtergram sequence. & 05:35:32, June 6, 2012. & & \ref{ssmcss}, \ref{psfdep1}\\
8 & A pair of filtergrams, one from each CCD, of similar bandpass ($-172\:{\rm m\AA{}}$ from line centre) and polarization (Stokes $I-V$), taken less than one minute apart of one another. & Around 05:36, June 6, 2012. & & \ref{psfdep2}\\
9 & The continuum filtergram recorded on each CCD whenever the SDO spacecraft passes orbital noon and midnight. A total of 1866 filtergrams from each CCD, from when the HMI commenced regular operation to the time of the study. & Around 06:00 and 18:00 daily, between May 1, 2010 and June 30, 2013. & & \ref{psfdep2}\\
\hline
\end{tabularx}
\end{table*}

When generating the various data products in the HMI data processing pipeline, filtergrams (from an interval of 1350-s for the 720-s data products and 270-s for the 45-s data products) are corrected for spatial distortion \citep{wachter12}, cosmic ray hits, polarization crosstalk \citep{schou12b} and solar rotation, and the filtergrams of similar polarizations averaged. These time-averaged filtergrams are then combined non-linearly to form the various data products \citep{couvidat12}. For the side CCD, these time-averaged filtergrams are outputted as the 720-s Stokes parameters product.

As a consequence of the correction for spatial distortion, the apparent PSF is different in the unprocessed and time-averaged filtergrams. The non-linearity of the algorithms used to derive the data products means they cannot be corrected for stray light by the deconvolution with the PSF. Instead, their restoration must go via restoring the unprocessed or time-averaged filtergrams.

The 249 continuum filtergrams considered (and all the other filtergram data utilised in the rest of the study) were corrected for spatial distortion. The retrieved PSF therefore represents stray light in undistorted HMI observations. This allows the generation of stray light-free data products through the deconvolution of the PSF from the time-averaged filtergrams. For the line-of-sight data products, this means correcting just the time-averaged Stokes $I+V$ and $I-V$ at each line position, a total of $2\times6=12$ images in each instance, instead of all the unprocessed filtergrams, which numbers 360 and 72 for the 720-s and 45-s data products respectively.

In this study we assumed Venus to be a perfect sphere \citep{archinal11}. Radial distance and azimuth are denoted $r$ and $\phi$, respectively. Azimuth is taken anti-clockwise from the CCD column axis such that zero is up.

\begin{figure}
\resizebox{\hsize}{!}{\includegraphics{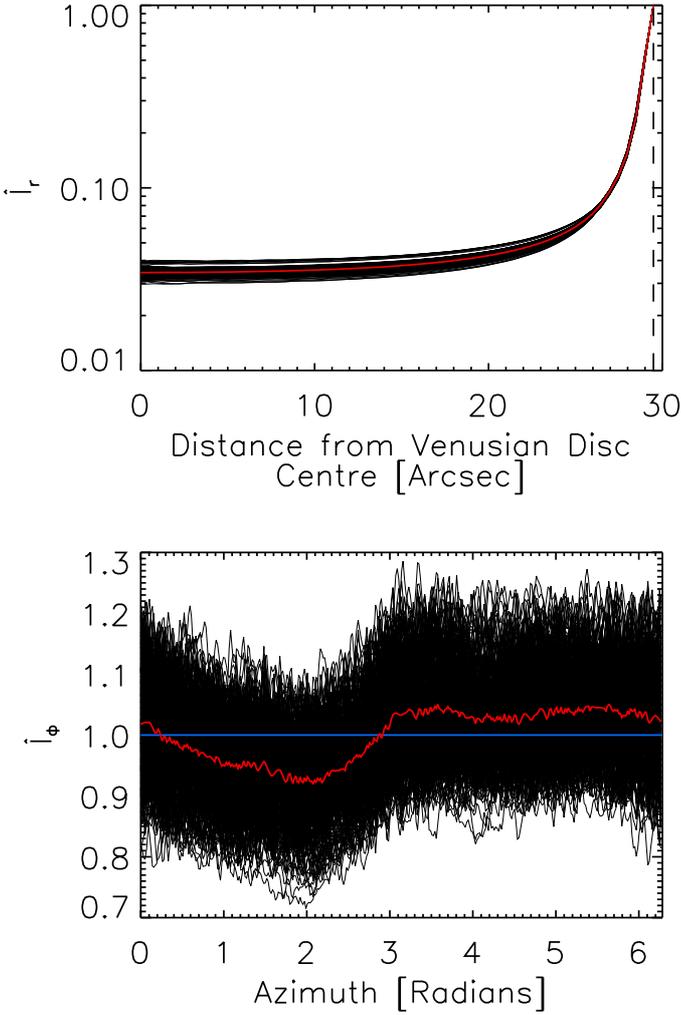}}
\caption{Intensity on the venusian disc in the 249 continuum filtergrams from the transit of Venus. Top: Intensity as a function of distance from the centre of the venusian disc, averaged over all azimuths and normalized to the level at the point of inflexion (dashed line), $\ir$. Bottom: Intensity versus azimuth along the edge of the venusian disc as given by the point of inflexion on $\ir$, normalized to the mean level, $\ia$. The red curves follow the mean $\ir$ and $\ia$ of all the filtergrams, while the blue line represents $\ia=1$.}
\label{stable2}
\end{figure}

The spatial distribution of intensity on the venusian disc, predominantly instrumental scattered light (aperture diffraction and stray light), varied significantly over the course of the transit (Fig. \ref{stable2}). The figure gives the intensity on the venusian disc in the 249 continuum filtergrams,
\begin{itemize}
	\item as a function of radial distance from the centre of the venusian disc, averaged over all azimuths and normalized to the level at the point of inflexion ($\ir$, top panel), and
	\item as a function of azimuth along the edge of the venusian disc as given by the point of inflexion on $\ir$, normalized to the mean level ($\ia$, bottom panel).
\end{itemize}
Also plotted are the mean $\ir$ and $\ia$ of all the filtergrams, $\irm$ and $\iam$ (red curves). The radius of the venusian disc as given by the point of inflexion on $\ir$ is, to 0.1 arcsec, constant at 29.5 arcsec. The fluctuation in the intensity on the venusian disc over the course of the transit arises from changes in the solar background from granulation, $p$-mode oscillations and limb darkening, as well as the variation of the PSF with position in the FOV.

\begin{figure}
\resizebox{\hsize}{!}{\includegraphics{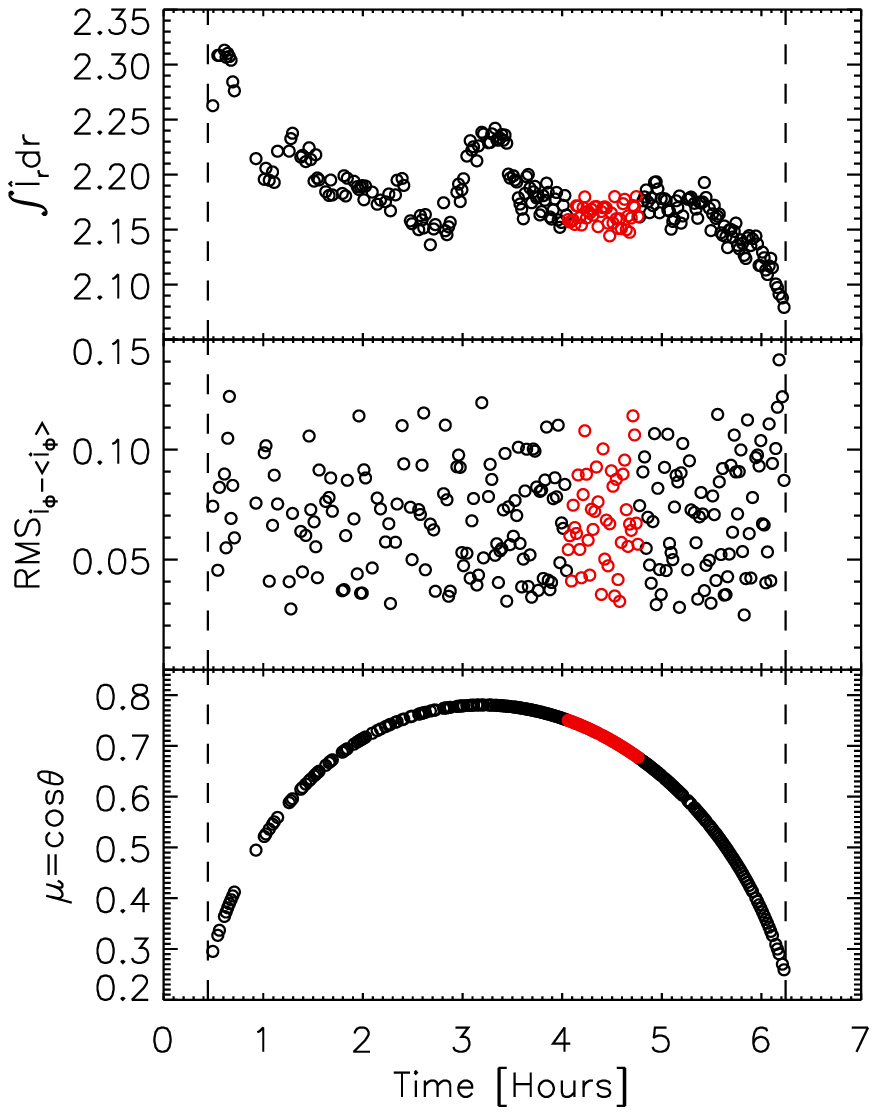}}
\caption{The integral under $\ir$, $\sir$ (top), the RMS difference between $\ia$ and $\iam$, $\sia$ (middle) and the position of the centre of the venusian disc in terms of the cosine of the heliocentric angle, $\mu=\cos\theta$ (bottom) as a function of time from 22:00:00 UTC, June 5, 2012. The red circles highlight the values for the filtergrams used to derive the PSF. The dashed lines mark the approximate time of the second and third contacts; the period between which the venusian disc was entirely within the solar disc from SDO's position.}
\label{stable1}
\end{figure}

To quantify the variation in the spatial distribution of intensity on the venusian disc over the course of the transit, and the influence of the changing solar background and the variation of the PSF with position in the FOV, we computed, for each of the 249 continuum filtergrams, the following two quantities.
\begin{itemize}
	\item The integral under $\ir$ from the centre of the venusian disc to the point of inflexion, $\sir$. The broader the PSF at a given position in the FOV, the brighter the venusian disc is relative to the level at its edge, and the greater this integral.
	\item The root-mean-square or RMS difference between $\ia$ and $\iam$, $\sia$. The closer the agreement in the trend with azimuth between $\ia$ and $\iam$, the smaller this quantity. $\sia$ reflects changes in the isotropy of the PSF (such as, from astigmatism and coma aberrations) and variation in the spatial distribution of intensity of the solar background.
\end{itemize}
$\sir$ and $\sia$ are plotted along the trajectory of the venusian disc, given in terms of the cosine of the heliocentric angle, $\mu=\cos\theta$ of the disc centre, as a function of time in Fig. \ref{stable1}.

$\sir$ changed slowly but notably over the course of the transit and the variation appears to be uncorrelated to distance from solar disc centre (and therefore, limb darkening). This suggests that the width of the PSF varies significantly along the path of Venus in the FOV. $\sia$ showed a marked point-to-point fluctuation but is otherwise relatively low ($\lesssim0.1$) and exhibits no obvious trend with time. This indicates that the azimuth dependence of $\ia$ and $\iam$ is mainly driven by a persistent PSF anisotropy along the path of Venus in the FOV rather than limb darkening (see also Sect. \ref{psfderivation}). The marked point-to-point fluctuation is likely from granulation and $p$-mode oscillations. This is supported by the comparatively smooth time variation of $\sir$, which is less sensitive to these small-scale intensity fluctuations in the solar background due to the averaging over all azimuths.

From the trend of $\sir$ and $\sia$ with time, we surmise that the width of the PSF varies significantly along the path of Venus in the FOV and while the anisotropy of the PSF is relatively stable, it is obscured by granulation and $p$-mode oscillations in individual filtergrams. Based on these considerations, we retained just the 42 filtergrams highlighted in Fig. \ref{stable1} (red circles), taken over a 40-minute period where $\sir$ is relatively stable, for the derivation of the PSF. The selected filtergrams were aligned by the venusian disc and the average, hereafter referred to as the mean transit image, taken. The objective is to derive an image of Venus in transit where the influence of granulation and $p$-mode oscillations is minimal by averaging filtergrams where the PSF at Venus' position in the FOV is fairly similar.

The 249 continuum filtergrams examined alternated between four polarizations (Stokes $I+Q$, $I-Q$, $I+U$ and $I-U$). We found no systematic differences in the spatial distribution of intensity on the venusian disc between filtergrams of different polarizations and so made no distinction between them here. We repeated the analysis described in this subsection on the filtergrams recorded at the six regular wavelength positions, on the front CCD during the transit of Venus. There are systematic differences (in $\sir$ and $\sia$) between different positions, even between positions at similar distance from but opposite sides of the line centre. This suggests spectral line changes from effects unrelated to the stray light behaviour of the instrument may exert an influence on intensity measured on the venusian disc. Specifically, Doppler shifts from the motion of the spacecraft (SDO is in a geosynchronous orbit) and the rotation of the Sun. For this reason we restricted ourselves to the continuum filtergrams from the side CCD.

\subsection{PSF derivation method}
\label{psfderivation}

A bivariate polynomial function was fitted to the mean transit image. We excluded the circular area of 50 arcsec radii centred on the venusian disc (about three times the area of the venusian disc) from the regression. The extrapolation of the surface fit over this excluded area represents an estimate of the intensity if Venus had been absent. We filled a circular area in the surface fit, corresponding to the venusian disc, with zeroes. We will refer to the result, essentially a model of the mean transit image in the absence of an atmosphere in Venus, aperture diffraction and stray light, as the artificial image.

The PSF was determined by minimizing the chi-square between the convolution of the artificial image with a guess PSF and the mean transit image, in the circular area of 50 arcsec radii centred on the venusian disc. For this we employed the implementation of the Levenberg-Marquardt algorithm (LMA) included in the IDL Astronomy User's Library, mpfit2dfun.pro.

Intensity in the circular area was sampled at equal intervals in the radial (0.504 arcsec, the pixel scale) and azimuthal (1/360 radians) dimensions. This is to give intensity measured at each radius from the centre of the venusian disc more equal weight in the LMA optimization. The result is a closer agreement between the convolution of the artificial image with the guess PSF and the mean transit image on convergence than achieved by comparing the circular area in the artificial and mean transit images directly.

We scaled the artificial image by a factor prior to convolution with the guess PSF. We allowed this factor and the radius of the disc of zeroes in the artificial image to be free parameters in the LMA optimization, taking an initial value of unity (i.e., no scaling) and 29.5 arcsec (the position of the point of inflexion on $\ir$). This is to minimize error from any misestimation of the surface fit to the mean transit image and the radius of the venusian disc\footnote{The point of inflexion on $\ir$ is not an accurate indication of the position of the edge of the venusian disc due to the influence of the venusian atmosphere.}.

Following the example of previous studies \citep{martinezpillet92,mathew07,mathew09}, we attempted to model the guess PSF as the linear sum of various combinations of Gaussian and Lorentzian functions. Except here we allowed the amplitude of each Gaussian and Lorentzian component to vary sinusoidally with azimuth to accommodate PSF anisotropy. We also tried to set the ideal diffraction-limited PSF as one of the components. The guess PSF we found to reproduce the measured intensity in the artificial image best, denoted $K$, is given by
\begin{equation}
K(r,\phi)=\sum^{5}_{i=1}\left[1+A_i{\rm cos}(u_i\phi+v_i)\right]w_i\left[\frac{1}{2\pi\sigma^2_i}{\rm exp}\left(-\frac{r^2}{2\sigma^2_i}\right)\right].
\label{aiso5g}
\end{equation}
That is, the linear combination of five Gaussian functions, with weight $w_i$ and standard deviation $\sigma_i$, the amplitudes of which vary sinusoidally with azimuth, with amplitude $A_i$, period of $2\pi{}u_{i}$ radians (where $u_{i}\in\mathbb{Z}$) and phase $v_i$.

Modelling the guess PSF as the linear combination of four Gaussian or three Gaussian and a Lorentzian, as done in the cited works, still reproduces measured intensity in the artificial image reasonably well. The retrieved PSFs are however, negative at parts (i.e., unphysical) from the LMA converging to solutions where $|A_i|>1$. And introducing additional sinusoidal terms to the azimuth dependence of each Gaussian and Lorentzian function did not alleviate this problem. The linear combination of five Gaussians appeared necessary to reach a physical solution while reproducing the measured intensity in the artificial image in both the radial and azimuthal dimensions.

The guess PSF was applied to the artificial image by evaluating $K$ (Eqn. \ref{aiso5g}) at pixel scale intervals (0.504 arcsec) on a $251\times251$ grid, the centre element representing the origin ($r=0$), and taking the convolution of the artificial image with the result. On convergence, the value at each grid element represents the integral of the PSF over the element. The retrieved PSF therefore describes the pixel integrated PSF. This was done, instead of filling the grid with the pixel integrated value of $K$, for a practical reason. When correcting HMI observations for stray light via deconvolution with the PSF, it is the pixel integrated PSF that is required.

Care was taken to repeat the LMA optimization, varying the initial value of the free parameters, to reduce the likelihood that the solution lies in a local chi-square minimum. To accommodate the requirement that $u_{i}\in\mathbb{Z}$, we executed the LMA optimization with no constraint on the value of $u_{i}$, rounded the retrieved $u_{i}$ to the nearest integer and repeated the process with these parameters fixed.

The PSF derivation method described here implicitly assumes there is no interaction between solar radiation and the venusian atmosphere. We will qualify this statement, and detail the adjustments made to the artificial image and the mean transit image to account for the influence of the venusian atmosphere on the retrieved PSF in Sect. \ref{venusatm}. The method also ignores motion blurring from the lateral movement of the venusian disc relative to the solar disc. The displacement of the venusian disc within the exposure time of the instrument is, on average, about 0.015 pixels and can therefore be neglected without significant loss of accuracy.

In the following, we denote the PSF retrieved as described above, a preliminary estimate of the stray light behaviour of the instrument, by $\psfone$. The retrieved value of the parameters of $\psfone$ are listed in Table. \ref{aiso5gp}. The best fit value of the scale factor applied to the artificial image is 1.0029, and the radius of the disc of zeroes, 29.29 arcsec. Though only a slight departure from the initial values (unity and 29.5 arcsec), this correction to the amplitude of the artificial image and the size of the disc of zeroes effected a marked improvement in the chi-square statistic.

\begin{table*}
\caption{Parameter values (and associated formal regression error) of the guess PSF retrieved neglecting ($\psfone$) and accounting for the influence of the venusian atmosphere ($\psftwo$).}
\label{aiso5gp}
\centering
\begin{tabular}{lcccccc}
\hline\hline
PSF & Gaussian component & $w_i$ & $\sigma_i$ [Arcsec] & $A_i$ & $u_i$ & $v_i$ [Radians] \\
\hline
$\psfone$ & $i=1$ & $0.641\pm0.002$ & $0.470\pm0.001$ & $0.131\pm0.002$ & 1 & $-1.85\pm0.02$ \\
& $i=2$ & $0.211\pm0.002$ & $1.155\pm0.008$ & $0.371\pm0.006$ & 1 & $2.62\pm0.01$ \\
& $i=3$ & $0.066\pm0.002$ & $2.09\pm0.02$ & $0.54\pm0.01$ & 2 & $-2.34\pm0.01$ \\
& $i=4$ & $0.0467\pm0.0005$ & $4.42\pm0.02$ & $0.781\pm0.006$ & 1 & $1.255\pm0.004$ \\
& $i=5$ & $0.035\pm0.004$ & $25.77\pm0.04$ & $0.115\pm0.001$ & 1 & $2.58\pm0.01$ \\
\hline
$\psftwo$ & $i=1$ & $0.747\pm0.001$ & $0.417$ & $0.164\pm0.002$ & 1 & $-2.22\pm0.01$ \\
& $i=2$ & $0.126\pm0.003$ & $1.45\pm0.01$ & $0.48\pm0.01$ & 1 & $2.36\pm0.01$ \\
& $i=3$ & $0.049\pm0.003$ & $2.10\pm0.02$ & $0.74\pm0.04$ & 2 & $-2.36\pm0.01$ \\
& $i=4$ & $0.0428\pm0.0004$ & $4.66\pm0.02$ & $0.776\pm0.007$ & 1 & $1.194\pm0.006$ \\
& $i=5$ & $0.035\pm0.004$ & $26.16\pm0.05$ & $0.122\pm0.002$ & 1 & $2.63\pm0.01$ \\
\hline
\end{tabular}
\tablefoot{The PSFs are given by the linear combination of five Gaussian functions (Eqn. \ref{aiso5g}), denoted by $i$, with weight $w_i$ and listed in ascending order by the standard deviation, $\sigma_i$. The amplitude of each Gaussian component modulates sinusoidally with azimuth, with amplitude $A_i$, period of $2\pi{}u_i$ radians (where $u_{i}\in\mathbb{Z}$) and phase $v_i$. There are no associated formal regression errors for $u_i$, and in the case of $\psftwo$, $\sigma_{i=1}$ as the value of these parameters were fixed in the LMA optimization (see text).}
\end{table*}

\begin{figure}
\resizebox{\hsize}{!}{\includegraphics{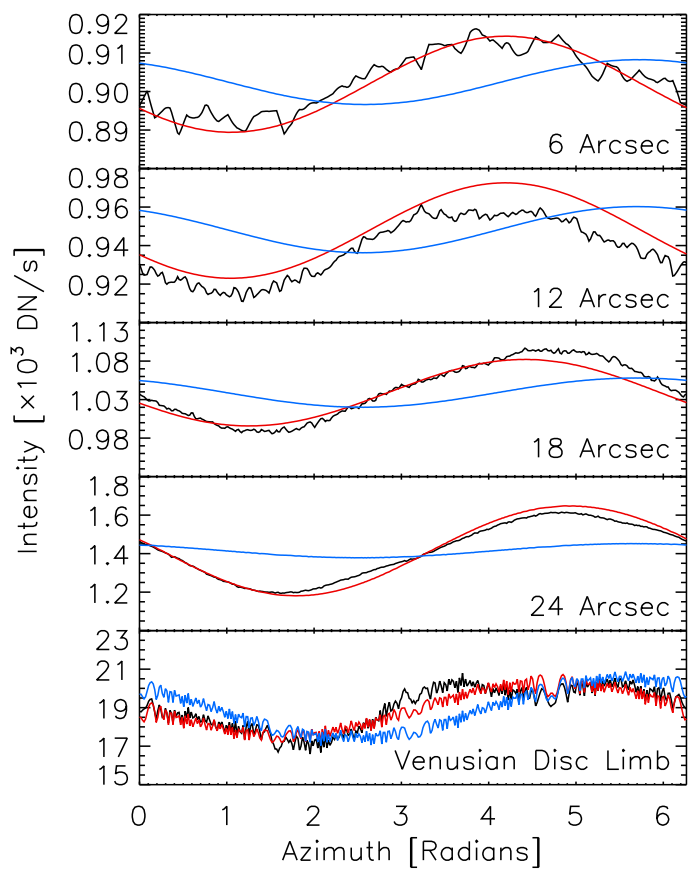}}
\caption{Intensity on the venusian disc at distances of 6, 12 18 and 24 arcsec from the centre, and along its limb (taking the radius of the venusian disc retrieved along with $\psfone$ by the LMA optimization, 29.29 arcsec), as a function of azimuth. The black curves follow the values from the mean transit image and the red curves the values reproduced in the artificial image by the convolution with $\psfone$. The blue curves represent the intensity obtained in the artificial image by fixing $A_i$ at zero. The intensity fluctuations along the venusian limb arises from aliasing artefacts.}
\label{aisofit1}
\end{figure}

\begin{figure}
\resizebox{\hsize}{!}{\includegraphics{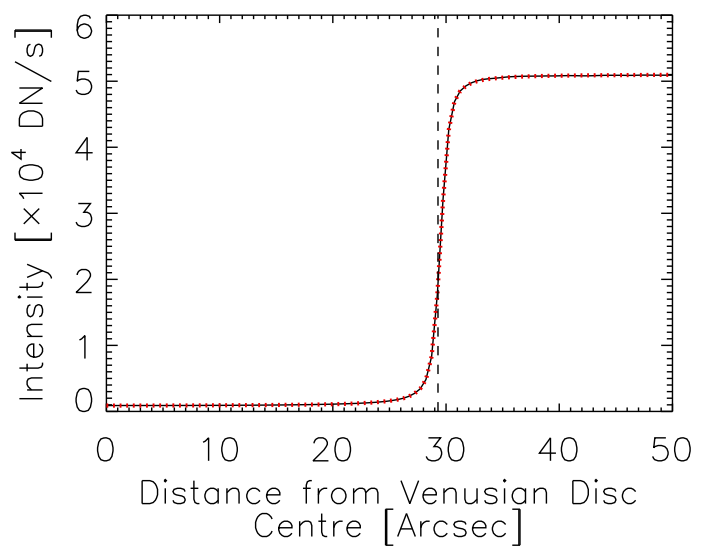}}
\caption{Radial intensity as a function of distance from the centre of the venusian disc; from the mean transit image (black solid curve) and the convolution of the artificial image with $\psfone$ (red dotted curve). The dashed line marks the position of the Venus limb as returned along with $\psfone$ by the LMA optimization.}
\label{aisofit2}
\end{figure}

In Fig. \ref{aisofit1} we plot the intensity along different radii from the centre of the venusian disc; from the mean transit image (black curves) and reproduced in the artificial image by the convolution with $\psfone$ (red curves). Also plotted is the intensity reproduced in the artificial image by fixing the $A_i$ at zero (blue curves). In this instance, the variation with azimuth arises solely from limb darkening, which enters the process through the surface fit to the mean transit image. Evidently, limb darkening alone cannot account for all the observed variation with azimuth, confirming that the PSF of the instrument is significantly anisotropic. By allowing the amplitude of each Gaussian component in the guess PSF to vary sinusoidally with azimuth, we are able to reproduce most of the observed intensity azimuth dependence. The close overall agreement between observed and reproduced intensities in the radial dimension is illustrated in Fig. \ref{aisofit2}.

Here we choose to describe the PSF as the linear combination of five Gaussian functions over more physically realistic models, such as the convolution of the diffraction-limited PSF with a Voigt function \citep{wedemeyerbohm08}. Apart from yielding a closer agreement between the PSF-blurred artificial image and the mean transit image, this functional form is more amenable to incorporating the complex azimuthal dependence. The retrieved parameter values of $\psfone$ (Table \ref{aiso5gp}) and the azimuth dependence of measured intensity on the venusian disc at different radii (Fig. \ref{aisofit1}) suggest that the overall amplitude and phase of the anisotropy of the PSF varies with radial distance.

The linear combination of Gaussian functions is not a physically realistic model of real PSFs for the following reasons:
\begin{itemize}
	\item It allows solutions with Strehl ratios exceeding unity, which is unphysical.
	\item The Fourier transform of the Gaussian function, and therefore the modulation transfer function (MTF) of such PSF models, is non-zero above the Nyquist limit. Correcting observations for stray light by the deconvolution with such a PSF can introduce aliasing artefacts from the enhancement of spatial frequencies above the Nyquist limit.
\end{itemize}
We will address these two potential issues in Sects. \ref{venusatm} and \ref{gc}, respectively.

The approach taken here to derive the PSF is broadly similar to that applied to images of Mercury in transit from Hinode/SOT/BFI by \cite{mathew09}. Specifically, by minimising the difference between observed intensity and that produced in a model of the aperture diffraction and stray light-free image (termed here the artificial image) by the convolution with a guess PSF. There are two significant departures.

Firstly, in this study, the artificial image is given by the surface fit to the mean transit image, with a disc of zeroes representing the venusian disc. In the cited work, the authors filled the mercurian disc in the recorded image with zeroes and took the result as the artificial image. As stated earlier in this subsection, having excluded the venusian disc and surroundings in the regression, the surface fit to the mean transit image is, in this excluded region, an approximation of the intensity had Venus been absent. For this we consider the approach taken here to yield a more realistic model of the instrumental scattered light-free image.

Secondly, as mentioned in the introduction, while the earlier effort assumed an isotropic form to the PSF, here we allowed the PSF to vary with azimuth. We were motivated by the observation that the stray light behaviour of the instrument is evidently anisotropic (Sect. \ref{dataselection} and Fig. \ref{aisofit1}).

\subsection{Image restoration method}
\label{imagerestoration}

To correct HMI observations for aperture diffraction and stray light, we utilised the implementation of the Richardson-Lucy algorithm, RLA \citep{richardson72,lucy74}, included in the IDL Astronomy User's Library, max\_likelihood.pro.

The RLA is an iterative method for restoring an image blurred by a known PSF, in our study, the guess PSF, $K$. Let $f_{k}$ denote the estimate of the restored image from the $k$-th iteration, $f_{k+1}$ is given by
\begin{equation}
f_{k+1}=f_{k}\circ\left(\left(f_{k}\ast{}K\right)\star{}K\right),
\label{rla}
\end{equation}
where the $\circ$, $\ast$ and $\star$ symbols represent the pixel-by-pixel product, convolution and correlation, respectively. The method has been shown, empirically, for data obeying Poisson statistics, to converge towards the maximum likelihood solution \citep{shepp82}. Following \cite{mathew09}, we employed a threshold for the chi-square between the original image and $f_{k}\ast{}K$ as the stopping rule. Here we set the threshold at $99.99\%$ confidence level.

\begin{figure*}
\centering
\includegraphics[width=17cm]{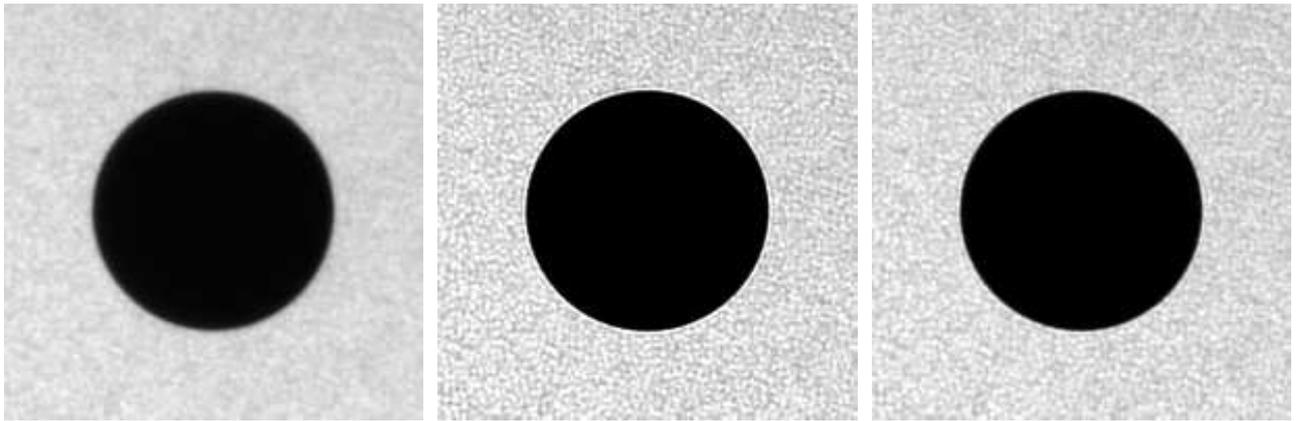}
\caption{$201\times201$ pixel inset, centred on Venus, of the test transit image; before (left) and after image restoration with $\psfone$ (middle), and with $\psftwo$ (right). The three grey scale plots are saturated at $6\times10^4\:\rm{DN/s}$, about $120\%$ of the mean photospheric level.}
\label{dim1}
\end{figure*}

\begin{figure*}
\centering
\includegraphics[width=17cm]{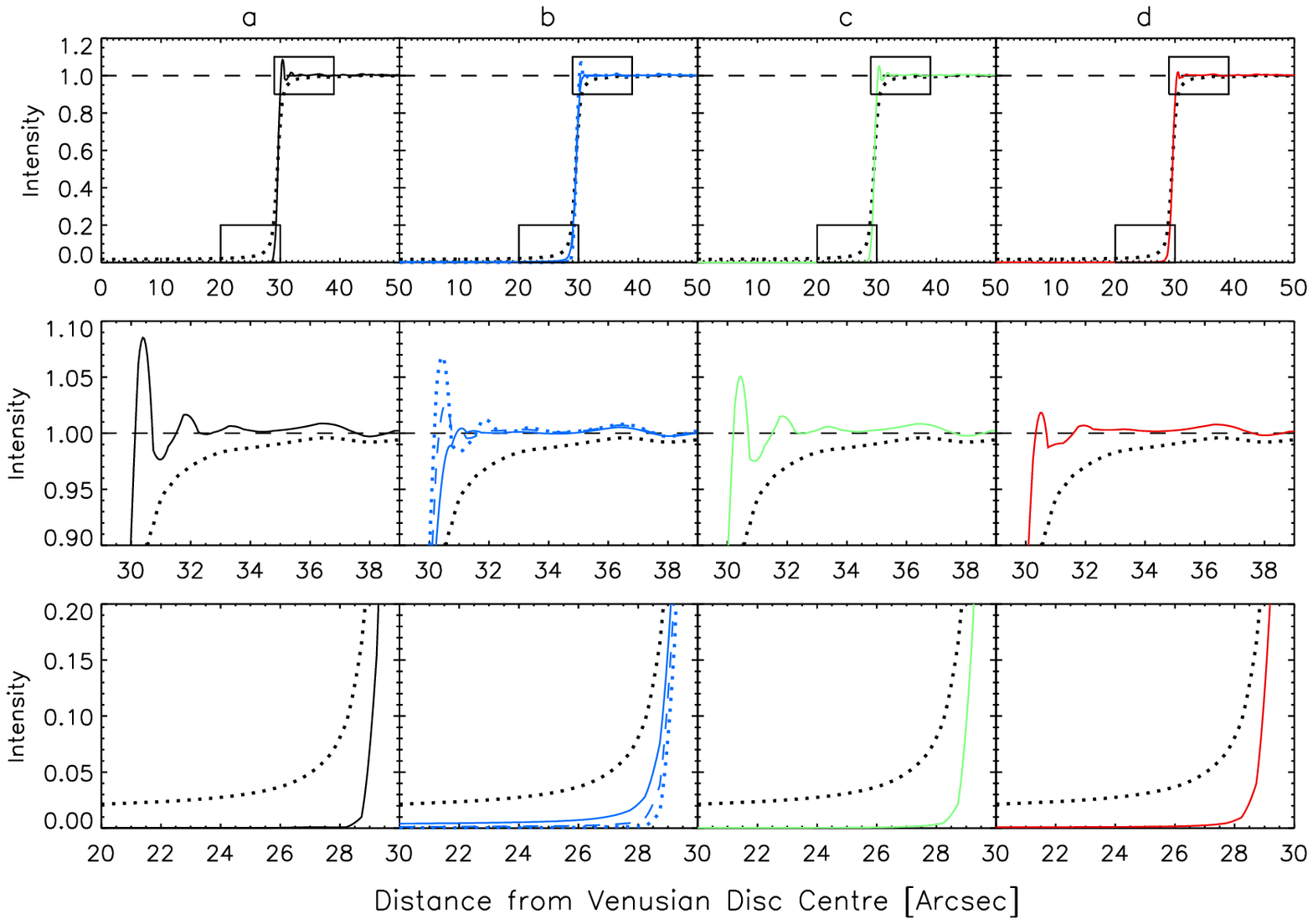}
\caption{Top: Radial intensity in the test transit image, corrected for limb darkening, as a function of distance from the centre of the venusian disc before (black dotted curves) and after image restoration with the various PSF estimates. a) With $\psfone$ (black solid curve). b) With the PSFs obtained by blurring the edge of the disc of zeroes in the artificial image with kernels representing Gaussian functions with standard deviations of 0.2, 0.3 and 0.4 arcsec (blue dotted, dashed and solid curves). c) With the PSF retrieved by subtracting the estimated aureole intensity from the mean transit image (green solid curve). d) With $\psftwo$ (red solid curve). Middle and bottom: Blow-up insets of the boxed areas. The radius of the venusian disc, a free parameter in the LMA optimization, is in all instances about 29.3 arcsec and not marked to avoid cluttering. The horizontal dashed lines follow unit intensity. The test transit image was corrected for limb darkening by normalizing it by the surface fit, computed as done for the mean transit image in Sect. \ref{psfderivation}.}
\label{dim2}
\end{figure*}

In Figs. \ref{dim1} and \ref{dim2}a we show the result of restoring one of the 42 continuum filtergrams averaged to yield the mean transit image (recorded at 02:25:37 UTC, June 6, 2012), hereafter referred to as the test transit image, with $\psfone$. The image restoration sharpened the test transit image considerably and removed most of the signal on the venusian disc. The restoration however, also left a ringing artefact around the venusian disc; manifest as the bright halo in the grey scale plot (middle panel, Fig. \ref{dim1}) and the peak in the radial intensity profile (black solid curve, Fig. \ref{dim2}a). Restoring other observations taken nearby in time (within a few hours of the test transit image), we found similar artefacts in the boundary of active region faculae, and sunspot penumbra and umbra.

\cite{mathew09} in the similar study with images of Mercury in transit from Hinode/SOT/BFI noted similar ringing artefacts around the mercurian disc upon image restoration with the RLA. The authors attributed it to Gibb's phenomenon, ringing artefacts in the Fourier series representation of discontinuous signals. In the PSF derivation and image restoration process described here, discrete Fourier transforms, DFTs were executed in convolution and correlation computations. We found that repeating the derivation of $\psfone$ and the restoration of the test transit image without performing any DFTs in the convolution and correlation computations had negligible effect on the ringing artefact, ruling out Gibb's phenomenon as the main cause in this instance. In the following subsection we will demonstrate the ringing artefact found here to arise from us not taking the interaction between solar radiation and the venusian atmosphere into account in deriving $\psfone$.

\begin{figure}
\resizebox{\hsize}{!}{\includegraphics{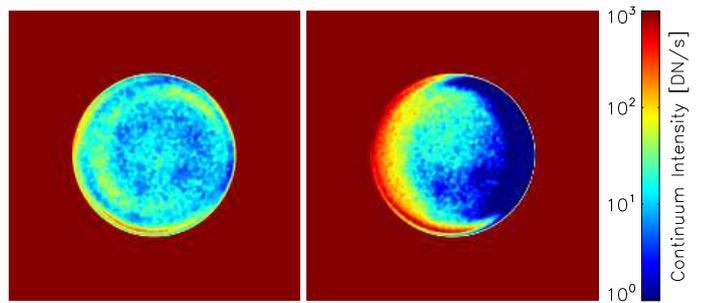}}
\caption{$201\times201$ pixel inset, centred on Venus, of the test transit image after image restoration with $\psfone$ (left) and the same, except with $A_{i}$ set to zero (right).}
\label{dim4}
\end{figure}

Figure \ref{dim4} is a colour scale plot of the venusian disc in the restored test transit image (left panel). The plot is saturated at about $2\%$ of the mean photospheric level to reveal the spatial distribution of residual intensity (instrumental scattered light not removed by the image restoration) on the venusian disc. Also shown is the result of restoring the test transit image with $\psfone$ excluding the anisotropy of the PSF by setting $A_{i}$ at zero (right panel). There is a gross, broadly east-west graduation of the residual intensity in the latter, not apparent in the former, where the residual intensity level is significantly more uniform across the venusian disc. This demonstrates the necessity to constrain the anisotropy of the PSF to properly correct HMI observations for instrumental scattered light.

\subsection{Interaction between solar radiation and the venusian atmosphere}
\label{venusatm}

As stated in Sect. \ref{psfderivation}, the PSF derivation method described so far builds on the assumption that there is no interaction between solar radiation and the venusian atmosphere.

In representing the venusian disc as a disc of zeroes in the artificial image, we have presumed that the body would, in the absence of aperture diffraction and stray light, be completely dark and exhibit a discrete edge. Diffusion and scattering of solar radiation in the venusian atmosphere can, however, render the edge of the venusian disc diffused.

The PSF is retrieved from matching the mean transit image to the convolution of the artificial image and the guess PSF. This is valid if all measured intensity came directly from the Sun. This is, however, not the case; there is a bright halo around the venusian disc when it is in transit (termed the aureole) from the refraction of solar radiation by the upper layers of the atmosphere towards the observer. 

\subsubsection{Diffusion and scattering of solar radiation in the venusian atmosphere}
\label{venusatm1}

To elucidate the influence of diffusion and scattering of solar radiation in the venusian atmosphere on the retrieved PSF we repeated the derivation, approximating the action of diffusion and scattering by blurring the edge of the disc of zeroes in the artificial image prior to the convolution with the guess PSF. We generated a copy of the artificial image that is unity everywhere outside the venusian disc and zero inside, convolved it with a Gaussian kernel, and scaled the original artificial image by the result. This procedure introduces Gaussian blur that is confined to near the edge of the disc of zeroes. We repeated the derivation of the PSF with different degrees of Gaussian blurring.

\begin{figure}
\resizebox{\hsize}{!}{\includegraphics{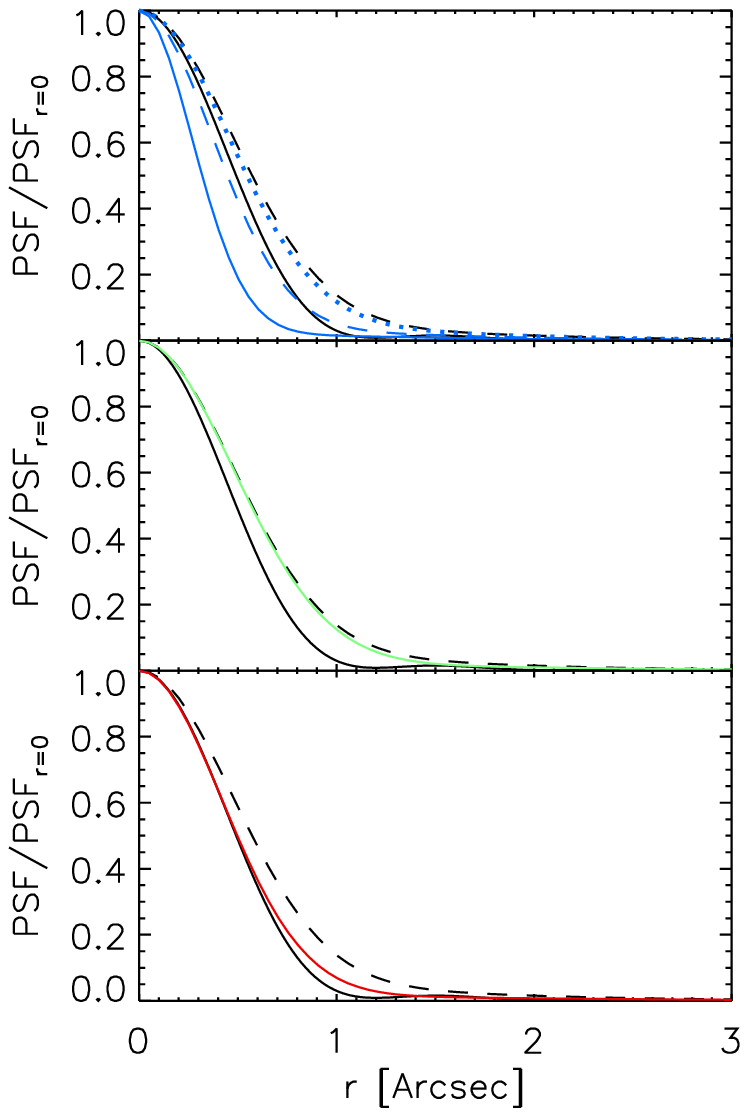}}
\caption{Comparison between $\psfone$ (black dashed curves), the ideal diffraction-limited PSF (black solid curves) and the other retrieved PSFs. The blue curves (top) represent the PSFs retrieved after blurring the edge of the disc of zeroes in the artificial image by kernels representing Gaussian functions with standard deviations of 0.2 (dotted), 0.3 (dashed) and 0.4 arcsec (solid). The green curve (middle) corresponds to the PSF found by subtracting the estimated aureole intensity from the mean transit image. The red curve (bottom) corresponds to $\psftwo$. For illustration purposes, we set the value of $A_{i}$ at zero (i.e., ignoring the azimuthal dependence) and normalized each PSF to the level at $r=0$. As the retrieved PSFs represent estimates of the pixel integrated true PSF, the ideal diffraction-limited PSF was smoothed with a box function of pixel scale width to allow a direct comparison.}
\label{dim3}
\end{figure}

In Fig. \ref{dim3} (top panel) we display the PSFs retrieved after blurring the edge of the disc of zeroes with kernels representing Gaussian functions with standard deviations of 0.2, 0.3 and 0.4 arcsec (blue dotted, dashed and solid curves) along $\psfone$ (black dashed curve). In Fig. \ref{dim2}b we have the radial intensity profile of the test transit image before and after image restoration with these PSFs. The stronger the blurring, the narrower the core of the PSF and the weaker the ringing artefact. The narrowest Gaussian kernel (0.2 arcsec) returned a PSF that is still very similar to $\psfone$ while the broadest (0.4 arcsec) yielded a PSF that is unphysical, significantly narrower at the core than the ideal diffraction-limited PSF. As the Gaussian blurring is confined to near the edge of the disc of zeroes, the retrieved PSFs do not differ significantly from $\psfone$ beyond a few arcseconds from the centre of the PSF.

\subsubsection{Refraction of solar radiation in the venusian atmosphere}
\label{venusatm2}

\begin{figure}
\resizebox{\hsize}{!}{\includegraphics{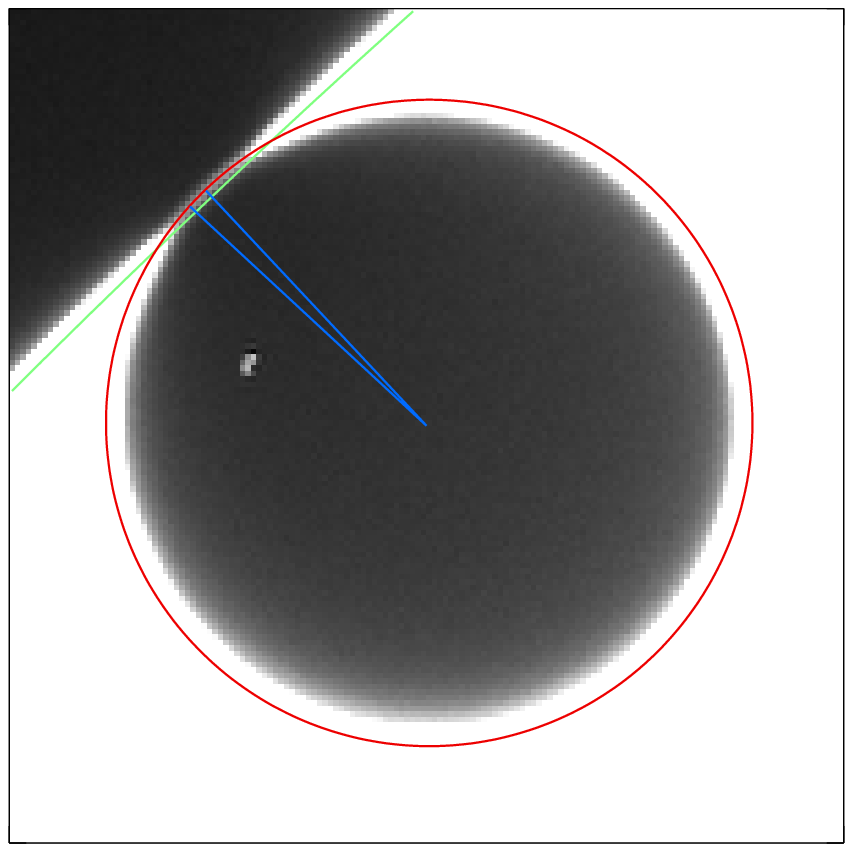}}
\caption{$151\times151$ pixel inset, centred on Venus, of the ingress image. The edge of the venusian disc (as given by the point of inflexion on $\ir$) and of the solar disc are indicated by the red and green contours respectively. They do not coincide with the apparent boundaries due to the low grey scale saturation level. The grey scale is saturated at $3000\:\rm{DN/s}$ ($\sim5\%$ of the mean photospheric level at disc centre) to allow the aureole, the bright arc on the part of the venusian disc outside the solar disc, which is much dimmer than the solar disc, to be visible. The blue lines mark the minor sector within which the intensity of the aureole is relatively stable with azimuth. The bright feature on the northwest quadrant of the venusian disc is an artefact of cosmic ray hits on the CCD.}
\label{aureole1}
\end{figure}

We estimated the contribution by the aureole to apparent intensity in the mean transit image. For this purpose we examined the $854\times854$ pixel crop, centred on the venusian disc, of a continuum filtergram taken shortly ($\sim10$ seconds) before the venusian disc moved completely into the solar disc (recorded at 22:25:33 UTC, June 5, 2012), hereafter referred to as the ingress image. The ingress image is expressed as a grey scale plot in Fig. \ref{aureole1}.

The aureole is only directly observable at ingress and egress (i.e., when the venusian disc is only partially within the solar disc), in the part of the venusian disc outside the solar disc. This is because the aureole is much dimmer than the photosphere and therefore difficult to distinguish from direct solar radiation. Generally, the intensity of the aureole increases with the proportion of the venusian disc sitting inside the solar disc \citep{tanga12}. Therefore, observations taken right before second contact (such as the ingress image) or right after third contact give the closest direct indication of the intensity of the aureole when the venusian disc is entirely within the solar disc. The intensity of the aureole also varies with azimuth. This is, at least in part, because it is modulated by the spatial distribution of photospheric intensity \citep{tanga12} and variation in the physical structure of the venusian atmosphere with latitude \citep{pasachoff11}.

Here we looked at the intensity of the aureole in the ingress image over the minor sector marked in Fig. \ref{aureole1} (blue lines), where it is relatively stable with azimuth.

\begin{figure}
\resizebox{\hsize}{!}{\includegraphics{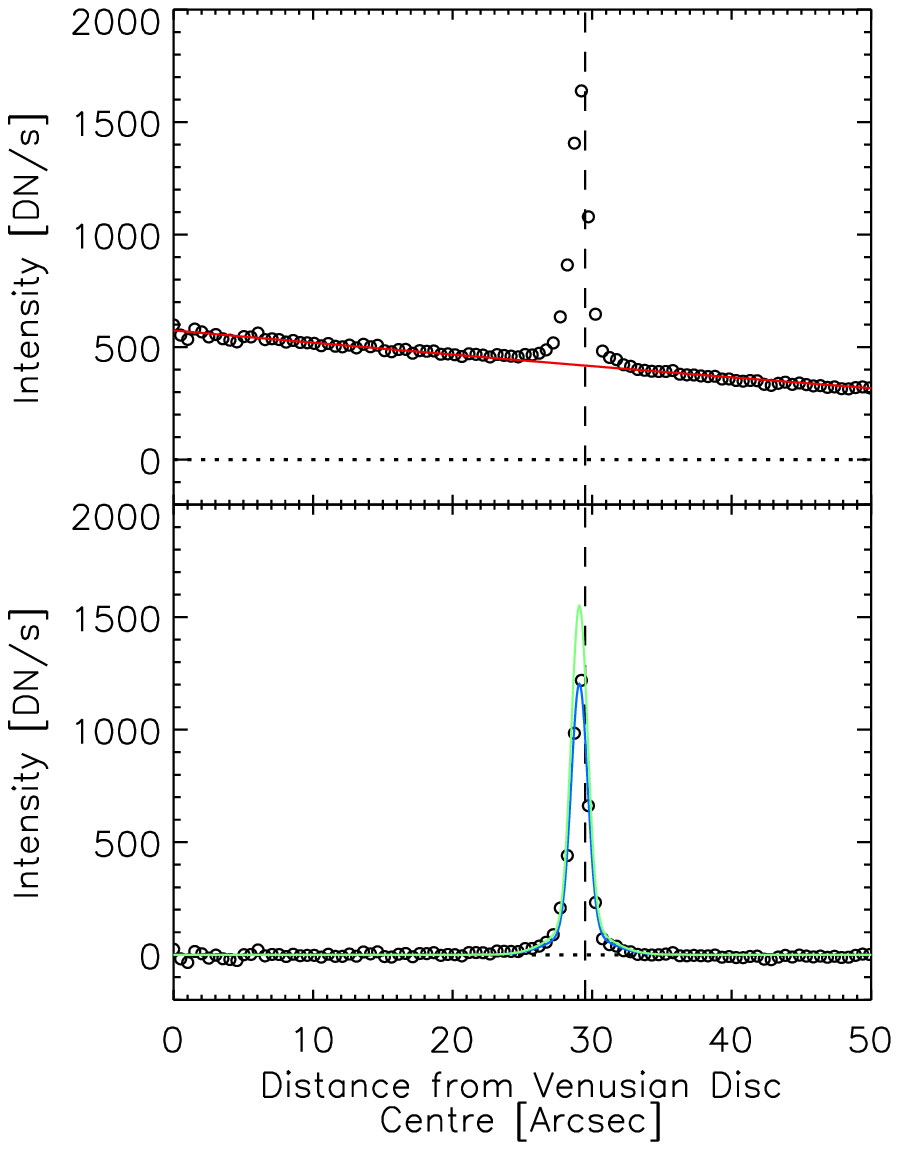}}
\caption{Intensity in the ingress image, averaged over the minor sector marked in Fig. \ref{aureole1}, as a function of distance from the centre of the venusian disc (circles); before (top panel) and after (bottom panel) subtracting the polynomial fit to the slowly varying background (red curve). The blue curve corresponds to the sum-of-two-Gaussians fit to the background-subtracted series, while the green curve is the same after scaling by the quotient of the total photospheric intensity behind the venusian disc in the mean transit image and in the ingress image. The black dashed and dotted lines denote the position of the Venus limb (as given by the point of inflexion on $\ir$) and the zero intensity level, respectively.}
\label{aureole2}
\end{figure}

The radial intensity profile over the minor sector marked in Fig. \ref{aureole1} is plotted in Fig. \ref{aureole2} (circles, top panel). The peak near the edge of the venusian disc (dashed line) corresponds to the aureole while the slowly varying background is largely instrumental scattered light from the solar disc. We subtracted the polynomial fit to the background (red curve) from the radial intensity profile. To the background-subtracted radial intensity profile (circles, bottom panel) we fit the linear combination of two Gaussian functions (blue curve). We then scaled this fit by the quotient of the integrated photospheric intensity behind the venusian disc in the mean transit image and in the ingress image\footnote{The intensity of the photosphere behind the venusian disc in the mean transit image is given by the surface fit described in Sect. \ref{psfderivation}. For the ingress image, we binned the image pixels on the solar disc by $\mu$, excluding the venusian disc and surroundings, and took the bin-averaged intensity. The intensity behind the venusian disc was then estimated from the polynomial fit to these bin-averaged intensities given the $\mu$ of each image pixel within the venusian disc.}. The result (green curve) represents an estimate of the radial intensity profile of the aureole in the mean transit image.

In Fig. \ref{dim3} (middle panel) we compare the PSF retrieved after first subtracting the estimated radial intensity profile of the aureole from the mean transit image (green curve) with $\psfone$ (black dashed curve). In Fig. \ref{dim2}c we have the radial intensity profile in the test transit image, before and after image restoration with this PSF. The effect of removing the contribution by the aureole to observed intensity on the retrieved PSF and the ringing artefact in the restored test transit image is similar as that from introducing Gaussian blur to the edge of the disc of zeroes in the artificial image. The retrieved PSF is slightly narrower than $\psfone$ at the core. The ringing artefact in the restored test transit image is slightly weaker. As the aureole is concentrated near the edge of the venusian disc, removing it from the mean transit image made little difference to the retrieved PSF beyond a few arcseconds from the core.

In removing the contribution of the aureole from the mean transit image as described above, we have made two simplifying assumptions:
\begin{itemize}
	\item One, that the intensity of the aureole is directly proportional to the integrated photospheric intensity behind the venusian disc.
	\item Two, that the intensity of the aureole does not change with azimuth.
\end{itemize}
\cite{tanga12} recently published a model of aureole intensity, relating it to the spatial distribution of photospheric intensity and physical structure of the venusian atmosphere. This is, to our knowledge, the only model of its kind reported in the literature. Given the fact that the aureole is blurred by instrumental scattered light, the uncertainties over the structure of the venusian atmosphere, and in the interest of simplicity, we favoured the rather approximate approach taken here over a more rigorous computation based on the model of \cite{tanga12}. The estimated peak intensity of the aureole in the mean transit image is about 1500 DN/s (blue curve, Fig. \ref{aureole2}), much smaller than the photospheric level ($\sim5\times10^4 \:\rm{DN/s}$, Fig. \ref{aisofit2}). Taking into account this as well as the relatively minor effect of subtracting the radial intensity profile of the aureole from the mean transit image on the retrieved PSF, we surmise that the uncertainty introduced by the two assumptions listed is likely minimal.

By repeating the derivation of the PSF and restoration of the test transit image, first blurring the edge of the disc of zeroes in the artificial image or subtracting the estimated contribution by the aureole to the mean transit image, we have demonstrated that the interaction between solar radiation and the venusian atmosphere has a palpable impact on the width of the core of the retrieved PSF. Both adjustments yielded PSFs that were narrower at the core compared to $\psfone$ (Fig. \ref{dim3}). And the narrower the core of the PSF, the weaker the ringing artefact in the restored test transit image (Fig. \ref{dim2}). The width of the core of $\psfone$, derived with no consideration of the venusian atmosphere, is over-estimated and the over-sharpening this produces when used to restore HMI data shows up as ringing artefacts near where the signal is changing rapidly.

\subsubsection{Final estimate of the PSF}

We arrived at our final estimate of the PSF by making the following changes to the derivation procedure described in Sect. \ref{psfderivation}.
\begin{itemize}
	\item Firstly, we subtracted the estimate of the radial intensity profile of the aureole from the mean transit image.
	\item Secondly, we blurred the edge of the disc of zeroes in the artificial image with a Gaussian kernel, the width of which we allowed to be a free parameter in the LMA optimization.
	\item Lastly, we fixed the width of the narrowest Gaussian component in the guess PSF such that the full width at half maximum, FWHM of the component is similar to that of the pixel-integrated ideal diffraction-limited PSF. The pixel integration was achieved by smoothing the ideal diffraction-limited PSF with a box filter of HMI pixel scale width.
\end{itemize}
The parameters of the PSF so derived, hereafter referred to as $\psftwo$, are summarized in Table \ref{aiso5gp}. The adjustments to the PSF derivation procedure yielded a PSF that is significantly narrower at the core compared to $\psfone$, though not more than the ideal diffraction-limited PSF (bottom panel, Fig. \ref{dim3}). The agreement between the aureole-subtracted mean transit image, and the convolution of the Gaussian-blurred artificial image with $\psftwo$ is similar as in the $\psfone$ instance, illustrated in Figs. \ref{aisofit1} and \ref{aisofit2}, and therefore not plotted here.

The best fit value of the scale factor applied to the artificial image is 1.0028 and the radius of the disc of zeroes 29.33 arcsec. The retrieved standard deviation of the Gaussian kernel is 0.26 arcsec. The scale height of the venusian atmosphere, at 15.9 km or approximately 0.08 arcsec, is of similar order. The degree of Gaussian blurring introduced is, as far as one can infer from such a comparison, physically plausible.

The intention here is to recover a conservative estimate of the PSF, making use of the fact that the PSF of the instrument cannot be narrower at the core than the ideal diffraction-limited PSF. Also, it was necessary to fix the width of the narrowest Gaussian component as allowing both this and the standard deviation of the Gaussian kernel to be free parameters leads to a degeneracy of the LMA optimization\footnote{Specifically, the LMA converged to different solutions for the PSF, some of which are narrower at the core than the ideal diffraction-limited PSF, depending on the initial value of the free parameters.}. Though a conservative estimate, restoring the test transit image with $\psftwo$ still removed most of the intensity on the venusian disc while largely suppressing the ringing artefacts (Figs. \ref{dim1} and \ref{dim2}d).

\begin{figure}
\resizebox{\hsize}{!}{\includegraphics{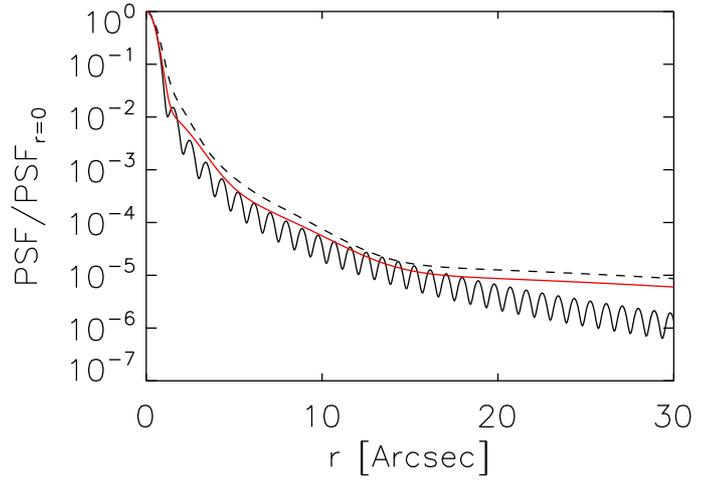}}
\caption{As in Fig. \ref{dim3} (bottom panel), but with the PSF on a logarithmic scale and over an extended radial range.}
\label{aisofit3}
\end{figure}

As stated in Sect. \ref{psfderivation}, a potential hazard of modelling the PSF as the linear combination of Gaussian functions is that it allows solutions with Strehl ratios greater than unity. This functional form is only appropriate when the weight and width of the broader Gaussian components, representing the non-ideal contribution to the PSF (instrumental effects other than aperture diffraction) are sufficiently high to avoid this \citep{wedemeyerbohm08}. As evident in Fig. \ref{aisofit3}, this is indeed the case here for both $\psfone$ (dashed curve) and $\psftwo$ (red curve). The greater integral under both PSFs compared to the pixel-integrated ideal diffraction-limited PSF, all normalized to the level at $r=0$, indicates Strehl ratios of less than unity. (We cannot compute the Strehl ratio of $\psfone$ and $\psftwo$ directly as they describe the pixel-integrated PSF.)

\subsection{Partial lunar eclipse and solar limb observations}
\label{psl}

As stated in the introduction, we consider the observations of Venus in transit the most appropriate available for recovering the PSF of the HMI, preferring them over data from partial lunar eclipses and of the solar limb.

It is challenging to constrain the PSF in both the radial and azimuthal dimensions with data from partial lunar eclipses due to the combination of the geometry, as well as jitter and defocus issues brought on by the lunar occultation itself.
\begin{itemize}
	\item The radius of curvature of the terminator, the edge of the lunar disc, is much greater than the width of the PSF. So at any given point along the terminator, the spatial distribution of signal smeared onto the lunar disc largely reflects the PSF in the direction of the centre of curvature. Therefore, a given partial lunar eclipse image only contains information about the PSF within a limited range of azimuths. Circumventing this limitation by looking at multiple images with the terminator at different orientations is not straightforward due to the variation of the PSF with position in the FOV.
	\item The terminator is uneven from lunar terrain. This makes it complicated to model the aperture diffraction and stray light-free image as we did here for the mean transit image with the artificial image (Sect. \ref{psfderivation}). A possible solution is to reduce the problem from 2D to 1D by looking at the radial intensity profile over segments of the terminator relatively free of lunar terrain features. This is, however, only appropriate when the terminator is near solar disc centre, where we can take the solar background to be uniform.
	\item The image stabilization system, ISS of the instrument was not always functional during the partial lunar eclipses due to the lunar disc blocking the diodes necessary for its operation, increasing the jitter.
	\item Having the lunar disc occult a significant proportion of the solar disc, and therefore greatly reducing the amount of impinging radiation, causes the front window of the instrument to cool, resulting in defocus \citep{schou12}.
	\item There are observations made when the lunar disc was just starting to cover the solar disc and not blocking the ISS diodes. These data do not suffer jitter and defocus problems, but we cannot resolve the lunar terrain issue by reducing the problem from 2D to 1D as described above, as the variation in the solar background from limb darkening is significant here.
\end{itemize}

Our approach in this study is to constraint the PSF by the spatial distribution of intensity about a closed bright and dark boundary as this allows us to recover the full azimuthal dependence (i.e., all directions). For this we can either employ observations of Venus in transit or of the solar limb. The PSF of HMI likely varies with position in the FOV, as shown for the part of the FOV transversed by Venus in Sect. \ref{dataselection}. The longer the boundary used to constrain the PSF, the greater the contribution by the variation of the PSF with position in the FOV to observed intensity fluctuation along and near the boundary, which introduces bias to the retrieved PSF. The venusian disc occupies only about $0.06\%$ of the FOV by area, and the solar disc, over $60\%$. The variation of the PSF over the part of the FOV occupied by the venusian disc is likely minimal, making these observations more suited for the purpose.

In view of the issues associated with deducing the PSF from partial lunar eclipse and solar limb data, we utilised the observations of Venus in transit though the interaction between solar radiation and the venusian atmosphere is challenging to account for, leaving us with only  a conservative estimate of the PSF (Sect. \ref{venusatm}).

\section{Application of the derived PSF to HMI observations}
\label{results}

\subsection{Granulation contrast}
\label{gc}

Restoring HMI data with $\psftwo$ is not exact. This is due to the approximate account of the influence of the venusian atmosphere in the derivation of $\psftwo$ (Sect. \ref{venusatm}) and from applying a single PSF to the entire FOV (so ignoring the variation of the PSF with position in the FOV, discussed in detail in Sect. \ref{psfdep1}). In this subsection we examine the effect of image restoration with $\psftwo$ on apparent granulation contrast, represented by the RMS intensity contrast of the quiet Sun\footnote{Intensity variation in the quiet Sun arises mainly from granulation.}. We compare the values deduced from HMI continuum observations and from synthetic intensity maps generated from a 3D MHD simulation. The purpose is to demonstrate that image restoration with $\psftwo$, with all its limitations, still yields reasonable estimates of the aperture diffraction and stray light-free intensity contrast.

The side CCD continued to observe in the continuum for about an hour after the end of the transit of Venus. For this analysis we employed a continuum filtergram from this period (recorded at 04:35:59 UTC, June 6, 2012), hereafter referred to as the test continuum filtergram. Of the various types of data available from HMI, the continuum filtergram represents the closest to a near instantaneous continuum capture (exposure time of $\sim0.135$ seconds). This implies minimal loss of apparent contrast from averaging in time. It is worth noting however, that the continuum bandpass ($-$344 m\AA{} from line centre), whilst close to the clean continuum, may be slightly affected by the far wing of the Fe I 6173 \AA{} line.

The intensity contrast of pixels corresponding to quiet Sun in the test continuum filtergram was computed largely following the method of \cite{yeo13}, who examined the intensity contrast of small-scale magnetic concentrations utilizing the 45-s continuum intensity and line depth data products from the front CCD. As in the cited work, the intensity contrast at a given image pixel is defined here as the normalized difference to the mean quiet-Sun intensity.

First, we identified magnetic activity present using the 45-s longitudinal magnetogram from the front CCD closest in time ($<1\:{\rm minute}$) to the test continuum filtergram\footnote{The 720-s longitudinal magnetogram data product from the side CCD, generated from the regular filtergram sequence, is evidently not available when this CCD is observing in the continuum.}. Let $\vbl/\mu$ denote the magnetogram signal, the mean line-of-sight magnetic flux density within a given image pixel, corrected (to first order) for foreshortening by the quotient with $\mu$. The magnetogram was resampled to register with the test continuum filtergram. Image pixels in the test continuum filtergram corresponding to points where $\bmu>10\:\rm{G}$ in the resampled magnetogram were taken to contain significant magnetic activity and masked, leaving quiet Sun.

The test continuum filtergram was corrected for limb darkening by normalizing it by a fifth order polynomial in $\mu$ fit to the quiet Sun pixels \citep[following][]{neckel94}. Let $\inorm$ denote the limb darkening corrected intensity. Next, we derived the mean $\inorm$ of the quiet Sun, $\iqs$ as a function of position on the solar disc. As similarly noted for the 45-s continuum and line depth data products by \cite{yeo13}, there are distortions in HMI filtergrams such that $\iqs$ is not at unity but varying with position on the solar disc. (This is not to be confused with the spatial distortion present in HMI data discussed in Sect. \ref{dataselection}.)

We sampled the solar disc at 16-pixel intervals in both the vertical and horizontal directions. At each sampled point, we retrieved the median intensity of all the quiet Sun pixels inside a $401\times401$ pixel window centred on the point. We then fit a bivariate polynomial surface to the values so obtained from the entire disc. This surface describes $\iqs$ as a function of position on the solar disc. The intensity contrast at a given image pixel is then given by the value of $\frac{\inorm}{\iqs}-1$ there.

\begin{figure}
\resizebox{\hsize}{!}{\includegraphics{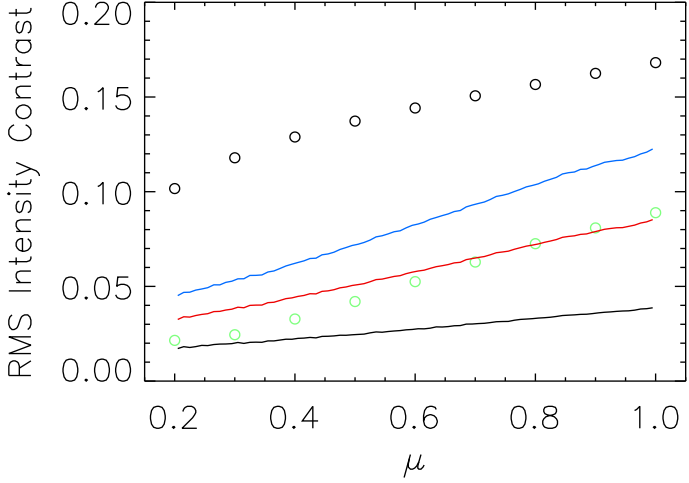}}
\caption{RMS intensity contrast of the quiet Sun in the test continuum filtergram as a function of $\mu$, before (black curve), and after image restoration with $\psfone$ (blue curve) and with $\psftwo$ (red curve). The black circles represent the values from the synthetic intensity maps and the green circles the same, rescaled to reflect the proportion arising from spatial frequencies up to the cutoff spatial frequency of the restored (with $\psftwo$) test continuum filtergram (see text).}
\label{contrastqs1}
\end{figure}

Finally, we derived the RMS intensity contrast of the quiet Sun as a function of $\mu$. To this end we grouped the quiet Sun pixels by $\mu$ in bins with a width of 0.01 and took the RMS intensity contrast within each bin. The results are expressed in Fig. \ref{contrastqs1} (black curve). Also plotted are the values from first restoring the test continuum filtergram with $\psfone$ (blue curve), and with $\psftwo$ (red curve). Quiet Sun pixels near the limb ($\mu<0.2$) were excluded. Towards the limb, the spread in measured intensity contrast in HMI data is dominated by scatter from the combination of the diminishing signal-to-noise ratio and the limb darkening correction \citep{yeo13}.

The decline in granulation contrast with distance from disc centre, seen here for the test continuum filtergram, is a known, well reported phenomenon \citep[see][and references therein]{sanchezcuberes00,sanchezcuberes03}. Also within expectation, image restoration resulted in greater RMS contrasts, by a factor of about 2.6 near limb in the $\psfone$ instance, going up to 3.2 at disc centre, and going from 1.9 to 2.2 for $\psftwo$.

The apparent granulation contrast is not only significantly enhanced by image restoration but also rather sensitive to differences between $\psfone$ and $\psftwo$. This makes the RMS intensity contrast of the quiet Sun a suitable check of the goodness of $\psftwo$ for the restoration of HMI data. We compared the values obtained here with values from synthetic intensity maps, artificial images of the quiet Sun produced from a numerical simulation. To this end, synthetic Stokes spectra were generated by applying the LTE radiative transfer package SPINOR/STOPRO \citep{solanki87,frutiger00}, to snapshots of a 3D MHD simulation performed with the MURaM code \citep{vogler03,vogler05}, as done in, for example, \cite{danilovic10,danilovic13}.

The simulation, set up as in \cite{danilovic13}, represents a layer encompassing the solar surface in the quiet Sun. The mean vertical magnetic flux density is $50\:{\rm G}$. The simulation ran over about 23 minutes solar time after reaching a statistically steady state. Synthetic Stokes profiles were computed for ten snapshots of the simulation output, recorded at intervals of approximately two minutes solar time. From each snapshot we produced nine synthetic intensity maps corresponding to $\mu$ of 0.2, 0.3, 0.4 and so on, up to 1.0, rotating the snapshot along one dimension. The computational domain of the simulation spans $6\times6\:{\rm Mm}$ in the horizontal, 1.4 Mm in depth, the top of the box lying about 0.5 Mm above the mean optical depth unity level, in a $288\times288\times100$ grid. This translates into a pixel scale of ($0.0287\mu$) arcsec and $0.0287$ arcsec, in the rotated and static direction, in the synthetic intensity maps.

The Stokes $I$ and $Q$ components of the synthetic spectra were convolved with a Gaussian function with a FWHM of 75 m\AA{} and sampled at $-$327 m\AA{} from the centre of the Fe I 6173 \AA{} line in order to yield synthetic intensity maps mimicking the polarization (Stokes $I+Q$) and bandpass of the test continuum filtergram. The FWHM and central wavelength of the continuum bandpass were estimated from the main lobe of the CCD centre filter transmission profile\footnote{The filter transmission profiles of the HMI varies slightly with position in the CCD. The central wavelength of the main lobe, at $-$327\AA{} from line centre, differs from the bandpass position of $-$344\AA{} stated earlier. Quoted bandpass positions for HMI are theoretical figures derived assuming the filter transmission profiles are delta functions.}.

The synthetic intensity maps were resampled in the foreshortened direction such that the pixel scale is similar along both dimensions. The RMS contrast of a given intensity map is given by the RMS value of $\frac{I}{\left\langle{I}\right\rangle}-1$ over all points, $\left\langle{I}\right\rangle$ denoting the mean intensity of the map. The mean RMS contrasts from the synthetic intensity maps at each $\mu$ level for which we simulated data are plotted along the measured values from the test continuum filtergram in Fig. \ref{contrastqs1} (black circles).

The RMS intensity contrast of the quiet Sun in the test continuum filtergram and the synthetic intensity maps cannot be compared directly due to the gross difference in the pixel scale (0.504 versus 0.0287 arcsec). Resampling the synthetic intensity maps to HMI's pixel scale is not feasible as the resampled synthetic intensity maps will extend only $16\times16$ pixels in the $\mu=1.0$ case, going down to $3\times16$ pixels for $\mu=0.2$. Simulations with considerably larger computational domains are necessary to yield synthetic intensity maps from which we can compute the RMS contrast at HMI's pixel scale with statistical confidence. What we did instead was to estimate, by comparing the power spectra of the synthetic intensity maps and the test continuum filtergram, the contribution to intensity variations in the synthetic intensity maps by spatial frequencies up to the resolution limit of the test continuum filtergram.

\begin{figure}
\resizebox{\hsize}{!}{\includegraphics{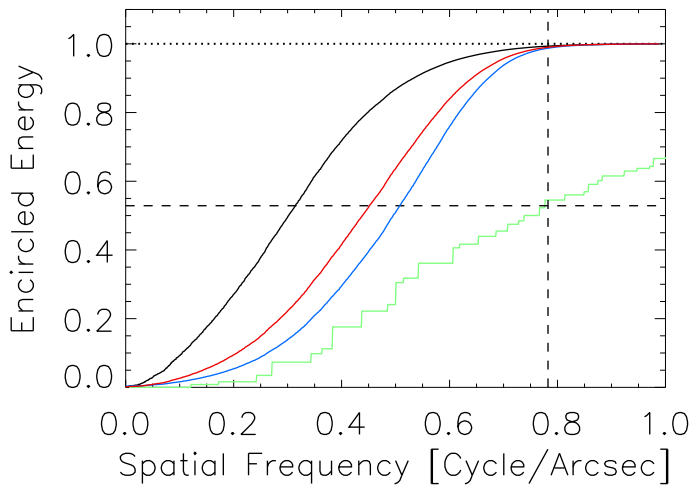}}
\caption{The encircled energy of the power spectrum of the $361\times361$ pixel inset, centred on the centre of the solar disc, of the test continuum filtergram, before (black), and after image restoration with $\psfone$ (blue) and with $\psftwo$ (red). The green series gives the encircled energy of the mean power spectrum of the synthetic intensity maps corresponding to $\mu=1.0$. The vertical dashed line marks the cutoff spatial frequency (see text) of the restored (with $\psftwo$) test continuum filtergram and the horizontal dashed line the encircled energy of the mean power spectrum of the $\mu=1.0$ synthetic intensity maps at this spatial frequency. The dotted line denotes encircled energy of unity.}
\label{analysisee}
\end{figure}

In Fig. \ref{analysisee} we plot the encircled energy of the power spectrum of the $361\times361$ pixel crop, centred on solar disc centre, of the test continuum filtergram. There are no sunspots present in this crop. We define the cutoff spatial frequency as the spatial frequency at which the encircled energy of the power spectrum reaches 0.99, taken here as an indication of the resolution limit. The cutoff spatial frequency is $0.75\:{\rm cycle/arcsec}$ for the original test continuum filtergram (black curve), and $0.79\:{\rm cycle/arcsec}$ and $0.78\:{\rm cycle/arcsec}$ for the iterations restored with $\psfone$ (blue curve) and with $\psftwo$ (red curve).

As mentioned in Sect. \ref{psfderivation}, image restoration with a PSF that is the linear combination of Gaussian functions can potentially introduce aliasing artefacts from the enhancement of spatial frequencies above the Nyquist limit ($0.99\:{\rm cycle/arcsec}$ for HMI). While image restoration with $\psfone$ and $\psftwo$ enhanced image contrast, indicated here by the rightward displacement of the encircled energy profile for the restored iterations of the test continuum filtergram, it made little difference to the resolution limit which is also significantly lower than the Nyquist limit. Even after image restoration, almost all energy is confined to spatial frequencies well below the Nyquist limit. Aliasing artefacts from the restoration, if present, are likely negligible.

At each $\mu$ level for which we generated synthetic intensity maps, we computed the power spectrum of each intensity map and then the encircled energy of the mean power spectrum. Following that we estimated the encircled energy at the spatial cutoff frequency of the restored (with $\psftwo$) test continuum filtergram, illustrated in Fig. \ref{analysisee} for $\mu=1.0$. The encircled energy here gives the proportion of observed intensity variation in the synthetic intensity maps arising from spatial frequencies up to the spatial cutoff frequency of the test continuum filtergram. The product of this quantity with the RMS intensity contrast of the quiet Sun from the synthetic intensity maps (green circles, Fig. \ref{contrastqs1}) then represents an approximation of the RMS contrast if the spatial resolution of the synthetic intensity maps were similar to that of the test continuum filtergram. This treatment is very approximate, ignoring the fact that the spatial frequency response of the test continuum filtergram and synthetic intensity maps, up to the cutoff, are in all likelihood not similar.

In this analysis we had,
\begin{itemize}
 \item defined the cutoff spatial frequency as the level where the encircled energy of the power spectrum reaches 0.99, and
 \item used the cutoff spatial frequency of the copy of the test continuum filtergram restored with $\psftwo$ to find the factors by which to rescale the RMS intensity contrast of the synthetic intensity maps (as the comparison between this restored version of the test continuum filtergram and the synthetic intensity maps is of greatest interest).
\end{itemize}
As stated above, the restoration of the test continuum filtergram made little difference to the cutoff spatial frequency. Also, the encircled energy of the power spectrum of the synthetic intensity maps does not vary strongly with spatial frequency in the regime of the cutoff spatial frequency of the test continuum filtergram, as visibly evident for the $\mu=1.0$ example in Fig. \ref{analysisee}. Hence, tests showed that the level of the rescaled RMS contrast is not sensitive to small variations in the threshold encircled energy level chosen in the definition of the cutoff spatial frequency. The result is also not changed significantly if we employ the cutoff spatial frequency of the unrestored and restored with $\psfone$ versions of the test continuum filtergram to derive the rescaling factors instead.

The RMS intensity contrast of the quiet Sun in the restored (with $\psftwo$) test continuum filtergram (red curve, Fig. \ref{contrastqs1}) and in the synthetic intensity maps, rescaled as described above (green circles), are of gratifyingly similar magnitude, less than or close to 0.01 apart at most $\mu$, especially near disc centre. They do, however, differ in that the latter exhibits a steeper decline with distance from disc centre. The diverging trend with decreasing $\mu$ is likely, at least in part, from
\begin{itemize}
	\item the approximate way of accounting for the influence of the venusian atmosphere in the derivation of $\psftwo$ (Sect. \ref{venusatm}),
	\item applying a single PSF to the entire FOV, so ignoring the variation of the PSF with position in the FOV (discussed in detail in Sect. \ref{psfdep1}),
	\item the difference in the spatial frequency response of the test continuum filtergram and the synthetic intensity maps,
	\item sensor noise and its centre-to-limb variation (CLV), and
	\item Doppler shift of the spectral line from the motion of SDO and the rotation of the Sun, which may produce small, $\mu$-dependent effects on apparent intensity in the continuum bandpass through the line wing.
\end{itemize}
The observation that these two series are close, even with these factors present, confers confidence that image restoration with $\psftwo$, though not exact, returns a reasonable approximation of the true aperture diffraction and stray light-free intensity contrast.

A quantitative comparison of the RMS intensity contrast of the quiet Sun presented here for HMI and other measurements reported in the literature would require taking into account instrumental differences such as the spatial resolution and bandpass, which is beyond the scope of this study. Due to HMI's limited spatial resolution, the RMS contrast, even after image restoration with $\psftwo$, remains below the values returned from spaceborne and balloon-borne (i.e., similarly seeing-free) observatories at finer spatial resolutions, namely Hinode \citep{danilovic08,mathew09,wedemeyerbohm09} and SUNRISE \citep{hirzberger10}. (Note though, that the divergence is also due in part to the different bandpass of the various instruments.)

\subsection{Effect of image restoration on the Dopplergram, longitudinal magnetogram, continuum intensity and line depth data products}
\label{ssmcss}

In this subsection we discuss the effect of image restoration with $\psftwo$ on the Dopplergram, longitudinal magnetogram, continuum intensity and line depth data products. We examine the influence on the apparent continuum and line-core intensity, and magnetic field strength of small-scale magnetic concentrations, as well as sunspots and pores. We will also describe the result of image restoration on the apparent amount of magnetic flux on the solar surface and the line-of-sight velocity.

For this purpose we utilised a set of simultaneous (generated from the same sequence of filtergrams) 720-s Dopplergram, longitudinal magnetogram, continuum intensity and line depth images from the side CCD, taken shortly after this CCD resumed collection of the regular filtergram sequence, about an hour after Venus left the solar disc (at 05:35:32 UTC, June 6, 2012). Here we will refer to the result of subtracting the line depth image from the continuum intensity image, giving the intensity in the Fe I 6173 \AA{} line, as the line-core intensity image.

\begin{figure*}
\centering
\includegraphics[width=17cm]{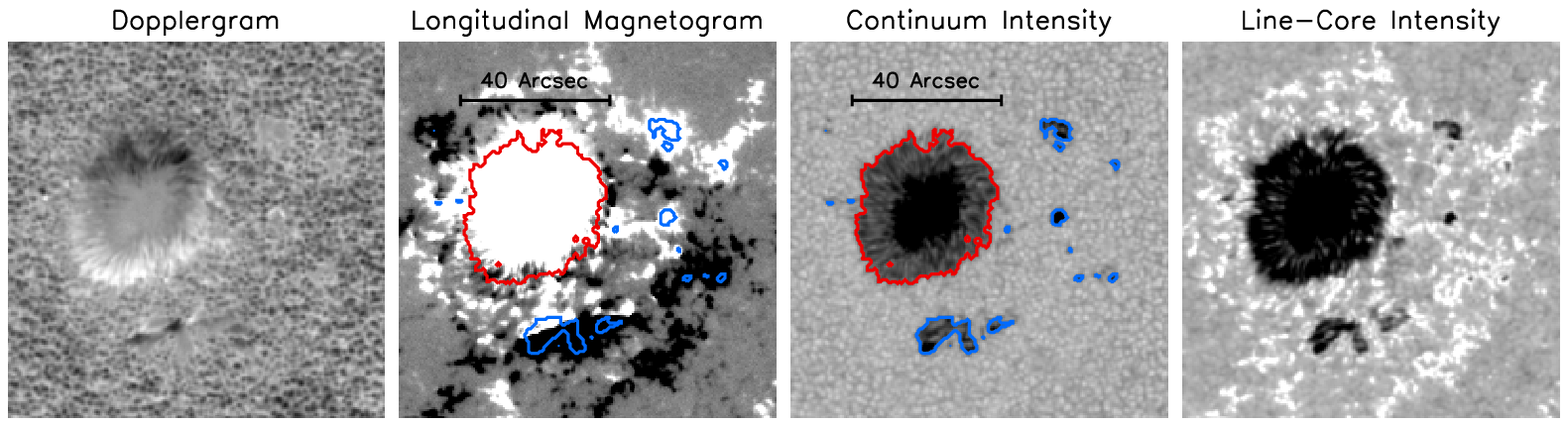}
\includegraphics[width=17cm]{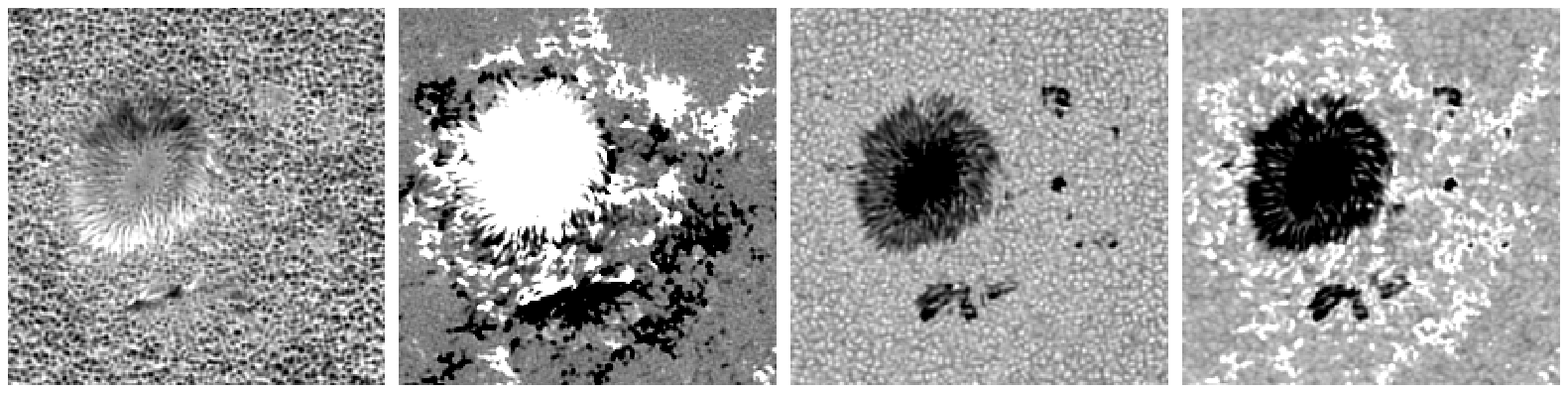}
\caption{$200\times200$ pixel ($101\times101$ arcsec) inset, near disc centre ($\mu>0.93$), encompassing active region NOAA 11494, of the simultaneously recorded 720-s data products examined, before (top) and after (bottom) image restoration with $\psftwo$. The grey scale is saturated at $-1200$ and $1200\:{\rm ms^{-1}}$ for the Dopplergrams, at $-100$ and $100\:{\rm G}$ for the longitudinal magnetograms, and at $0.6$ and $1.2$ for the continuum and line-core intensity images. The Dopplergram was corrected for the velocity of SDO relative to the Sun and for differential rotation (Sect. \ref{ssmcss4}). Both the continuum and line-core intensity images were normalized to the mean quiet-Sun level (Sect. \ref{ssmcss1}). The red and blue contours in the grey scale plot of the uncorrected longitudinal magnetogram and continuum intensity image follow $\inorm=\iqp$, the quiet Sun to penumbra boundary. The colour coding is to distinguish the big sunspot feature (red) from the smaller sunspots and pores (blue), treated separately in Fig. \ref{speg}.}
\label{spotinset}
\end{figure*}

As mentioned in Sect. \ref{dataselection}, HMI data products cannot be corrected for stray light by the deconvolution with the PSF but instead we must correct either the Stokes parameters or the filtergrams used to compute the data products. The data set was restored for stray light by applying image restoration with $\psftwo$ to the corresponding 720-s Stokes parameters, and returning the result to the HMI data processing pipeline. A $200\times200$ pixel inset of the original and restored version of the data set, near disc centre ($\mu>0.93$), featuring active region NOAA 11494, is shown in Fig. \ref{spotinset}. The enhanced image contrast and visibility of small-scale structures is clearly evident.

The 720-s Milne-Eddington inversion data product includes the vector magnetogram. Since the inversion procedure employed to obtain this data product assumes a magnetic filling factor of unity everywhere, the process treats noise in the Stokes $Q$, $U$ and $V$ parameters as signal, creating pixel-averaged horizontal magnetic field strengths of $\sim100\:{\rm G}$ in the vector magnetogram even in the very quiet Sun. For ease of interpretation we confined ourselves to the longitudinal magnetogram data product here.

\subsubsection{Intensity contrast and magnetogram signal of small-scale magnetic concentrations}
\label{ssmcss1}

Both the original and restored continuum and line-core intensity images were normalized by the fifth order polynomial in $\mu$ fit to the quiet Sun pixels. Then the intensity contrast at each image pixel was computed following the procedure applied to the test continuum filtergram in Sect. \ref{gc}. For the line-core intensity image, the normalization not only corrects for limb darkening, but also the centre-to-limb weakening of the Fe I 6173\AA{} line \citep{norton06,yeo13}.

Sunspots were identified by applying a continuum intensity threshold representing the quiet Sun-to-penumbra boundary, denoted $\iqp$. We took the threshold value for MDI continuum intensity images, taken at 6768 \AA{}, from \cite{ball12}, 0.89, and estimated the equivalent level at HMI's wavelength, 6173 \AA{}. Assuming sunspots to be perfect blackbodies and an effective temperature of $5800\:{\rm K}$ for the quiet Sun, the result is a threshold value of 0.88. This is a crude approximation, ignoring the difference in spatial resolution and variation in the continuum formation height with wavelength \citep{solanki98,sutterlin99,norton06}. Pores were also isolated by the application of this threshold. In the following we count these features to the sunspots and do not mention them separately.

\begin{figure}
\resizebox{\hsize}{!}{\includegraphics{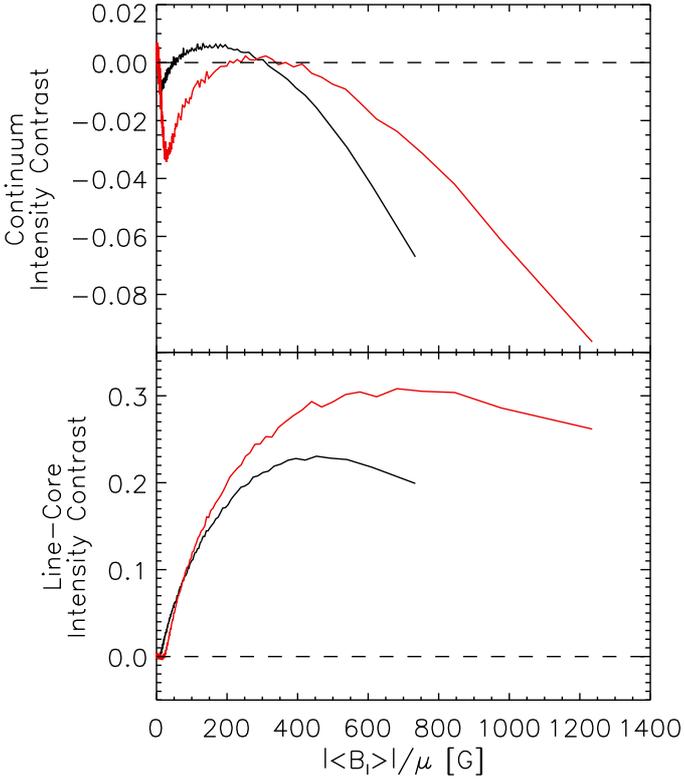}}
\caption{The continuum (top) and line core (bottom) intensity contrast of small-scale magnetic concentrations, near disc centre ($\mu>0.94$), as a function of $\bmu$. The black and red curves correspond to the values from the original and restored (with $\psftwo$) data sets, respectively.}
\label{contrastnf}
\end{figure}

We selected the image pixels where $\mu>0.94$ (i.e., near disc centre), excluding sunspots (i.e., all points with $\inorm<\iqp$) and all points within three pixels of a sunspot. The selected points were binned by $\bmu$ such that we end up with 800 bins of equal population. We then took the mean $\bmu$, as well as the median continuum and line-core intensity contrast within each bin. The values for the uncorrected and restored copy of the data set are represented by the black and red curves respectively in Fig. \ref{contrastnf}. These profiles depict the intensity contrast of small-scale magnetic concentrations as a function of $\bmu$, which serves as an approximate proxy of the magnetic filling factor \citep{ortiz02,yeo13}, in the continuum and core of the Fe I 6173 \AA{} line. This is effectively a repeat of part of the analysis of \cite{yeo13}, except now on HMI observations corrected for aperture diffraction and stray light.

Comparing the magnetogram with the continuum intensity image, we found the magnetogram signal associated with sunspots to extend beyond the continuum intensity boundary, given by the $\inorm=\iqp$ locus (for example, in Fig. \ref{spotinset}). This was similarly noted by \cite{yeo13}, who attributed it to the encroachment of the magnetic canopy of sunspots, and the smearing of polarized signal originating from sunspots onto its surrounds by instrumental scattered light. Hence, close to sunspots, the magnetogram signal is not entirely from local, non-sunspot magnetic features alone and would introduce a bias into the intensity contrast versus $\bmu$ profiles (Fig. \ref{contrastnf}) if left unaccounted for. In the cited work pixels contiguous to sunspots and with $\bmu$ above a certain threshold level were masked. Here, we excluded only all points within three pixels of each sunspot, observing that excluding pixels further than this distance made no appreciable difference to the resulting intensity contrast versus $\bmu$ profiles. This measure is sufficient here as, unlike in the earlier study which examined almost all disc positions ($\mu>0.1$), we are only looking at image pixels near disc centre ($\mu>0.94$). Near disc centre, the influence of magnetic canopies, which are largely horizontal, on the longitudinal magnetogram signal near sunspots is not as significant or extensive as at disc positions closer to the limb.

The continuum and line-core intensity contrast versus $\bmu$ profile of small-scale magnetic concentrations near disc centre presented here for the uncorrected data set (black curves, Fig. \ref{contrastnf}) is nearly identical to that by \cite{yeo13} (Figs. 9 and 10 in their paper), who employed similar data and method of derivation. The profiles from the restored data set (red curves) span a wider range, by a factor of about 1.3 in the continuum and line-core intensity contrast, and 1.7 in $\bmu$, but are qualitatively similar in form.

Image restoration produced an absolute increase in the continuum and line-core intensity contrast everywhere except around the peak of the continuum contrast versus $\bmu$ profile. The lower maximum in the profile from the restored data set, compared to the profile from the uncorrected data set ($2.6\times10^{-3}$ versus $5.7\times10^{-3}$), is likely from the enhanced contrast of dark intergranular lanes.

\begin{figure}
\resizebox{\hsize}{!}{\includegraphics{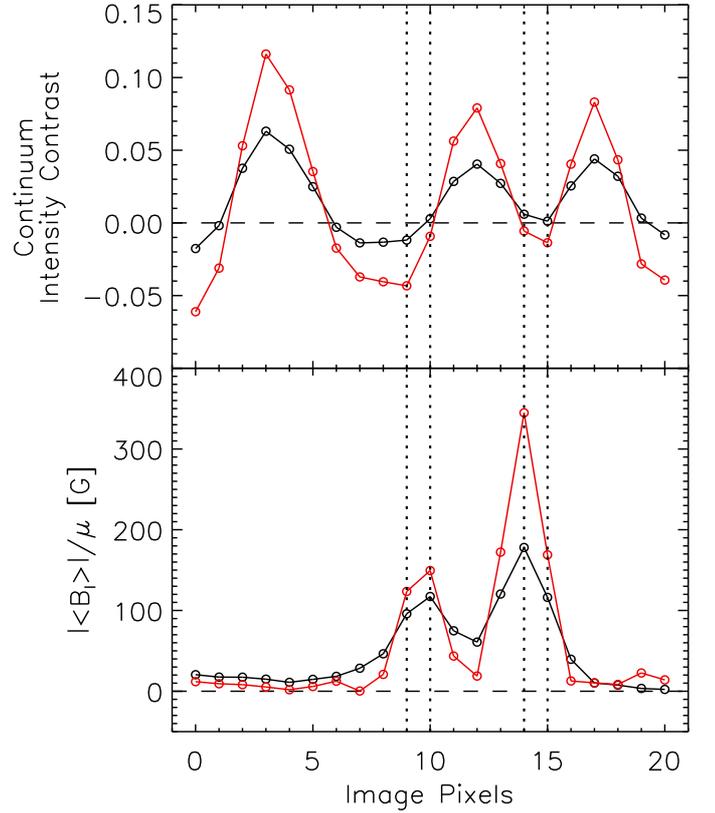}}
\caption{The uncorrected (black circles) and restored (with $\psftwo$, red circles) continuum intensity contrast (top) and $\bmu$ (bottom) along a 21-pixel section across magnetic features near disc centre ($\mu=0.97$). The plotted points represent image pixel values and are connected by straight lines to aid the eye. The dotted lines highlight the image pixels inside the magnetic features where the stray light correction effected a decrease in continuum intensity contrast accompanied by an increase in $\bmu$. The dashed lines follow the zero level.}
\label{contrastnfpeak}
\end{figure}

In Fig. \ref{contrastnfpeak} we show the uncorrected and restored continuum intensity and $\bmu$  along a 21-pixel cut across example magnetic features near disc centre ($\mu=0.97$). The troughs and peaks in the intensity curve (top panel) correspond to intergranular lanes and granules respectively. The magnetic features, the peaks in the $\bmu$ curve (bottom panel), sit inside the intergranular lanes. The $\bmu$ level at the core of these magnetic features lie in the regime of the peak of the continuum contrast versus $\bmu$ profile. The stray light correction boosted the magnetogram signal at the core of these magnetic features but also rendered them darker here, even from positive contrast to negative, as radiation originating from nearby granulation is removed from the intergranular lanes. The spatial resolution of HMI is insufficient to resolve many of the magnetic elements. Consequently, measured intensities contain contributions not only from magnetic features but also from the intergranular lanes that host them.

The intensity contrast of small-scale magnetic concentrations in both the continuum and spectral lines component of the solar spectrum, in particular the variation with position on the solar disc and magnetic field strength, is an important consideration in understanding the contribution by these features to variation in solar irradiance \citep{yeo13}. To the extent tested, image restoration with $\psftwo$ enhanced measured intensity contrast and $\bmu$ significantly but made little qualitative difference to the dependence of apparent contrast on $\bmu$. The analysis here was restricted to image pixels near disc centre ($\mu>0.94$). To extend the analysis to other disc positions, we would need to examine multiple full-disc images from different times featuring active regions at various disc positions as done by \cite{ortiz02} and \cite{yeo13}, beyond the scope of this paper.

\subsubsection{Intensity and magnetogram signal of sunspots and pores}
\label{ssmcss2}

\begin{figure*}
\centering
\includegraphics[width=17cm]{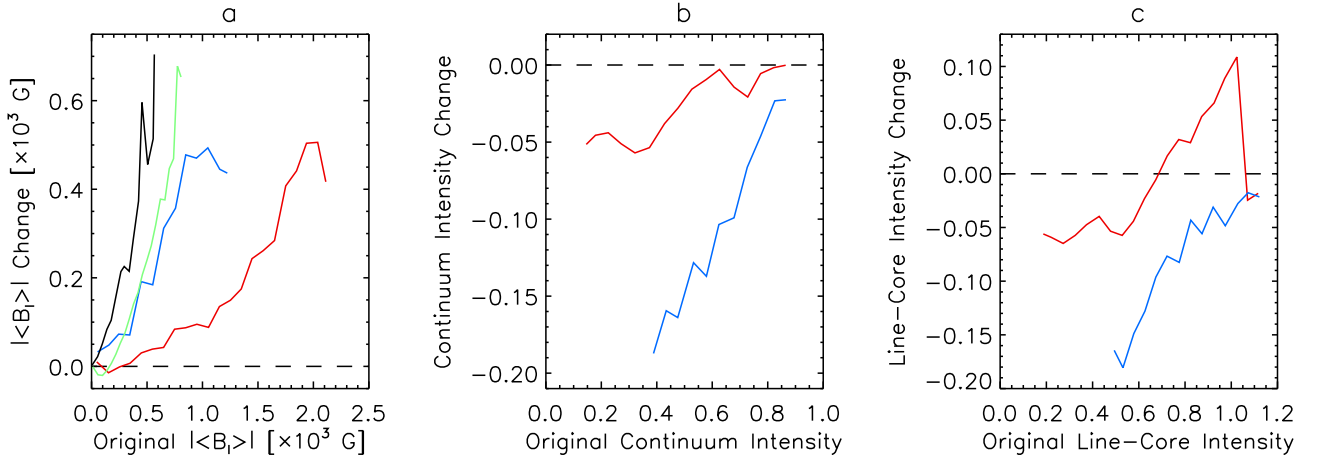}
\caption{The change in $\bl$ (left), as well as in the continuum (middle) and line-core intensity (right) introduced by image restoration with $\psftwo$, as a function of the original value. Both the continuum and line core intensities are normalized to the quiet-Sun level. The red and blue series follow the values derived from the sunspot and pore features encircled by the similarly colour-coded contours in Fig. \ref{spotinset}. The black and green curves (left) correspond to the change in $\bl$ in the quiet Sun field depicted in Fig. \ref{delta4}, and in a $201\times201$ pixel active region field near disc centre ($\mu>0.92$), respectively. The dashed lines mark the zero level.}
\label{speg}
\end{figure*}

In Fig. \ref{speg} we illustrate the change introduced by image restoration with $\psftwo$ on $\bl$, as well as the continuum and line-core intensity of the sunspots and pores defined by the $\inorm=\iqp$ contours in Fig. \ref{spotinset}. Signal enhancement is expressed as a function of the original level, separately for the big sunspot bounded by the red contours, and the smaller sunspots and pores bounded by the blue contours. We binned the image pixels by the uncorrected $\bl$ in intervals of $100\:{\rm G}$ and plotted the bin-averaged change in $\bl$ against the bin-averaged original $\bl$ (Fig. \ref{speg}a). This was repeated for the continuum and line-core intensity (Figs. \ref{speg}b and \ref{speg}c), taking a bin size of 0.05 in both instances.

Within expectation, the influence of image restoration on $\bl$ and intensity is highly correlated to the original values of these quantities. This comes largely from the fact that the darker regions, where $\bl$ is also typically higher, are more affected by stray light as scattered radiation forms a greater proportion of measured intensity, and therefore respond more strongly to image restoration. Also within expectation, the effect of restoration is more pronounced (greater absolute change) for smaller features, which are more susceptible to instrumental scattered light. An exception is the peak in the line-core intensity profile for the big sunspot feature (red curve, Fig. \ref{speg}c), which arose from the enhanced brightness of the bright filaments in the penumbra from the restoration, visible in Fig. \ref{spotinset}.

Given the variation in the response of sunspots to image restoration, it could have a profound effect on the apparent radiant and magnetic properties of these features. A full account of the effect of image restoration on sunspots, including the variation with size and disc position, would require examining a much larger sample of sunspots from multiple images taken at different times, which is outside the scope of this work (see \cite{mathew07} for such a study, based on MDI data).

\subsubsection{Amount of magnetic flux on the solar surface}
\label{ssmcss3}

\begin{figure*}
\centering
\includegraphics[width=17cm]{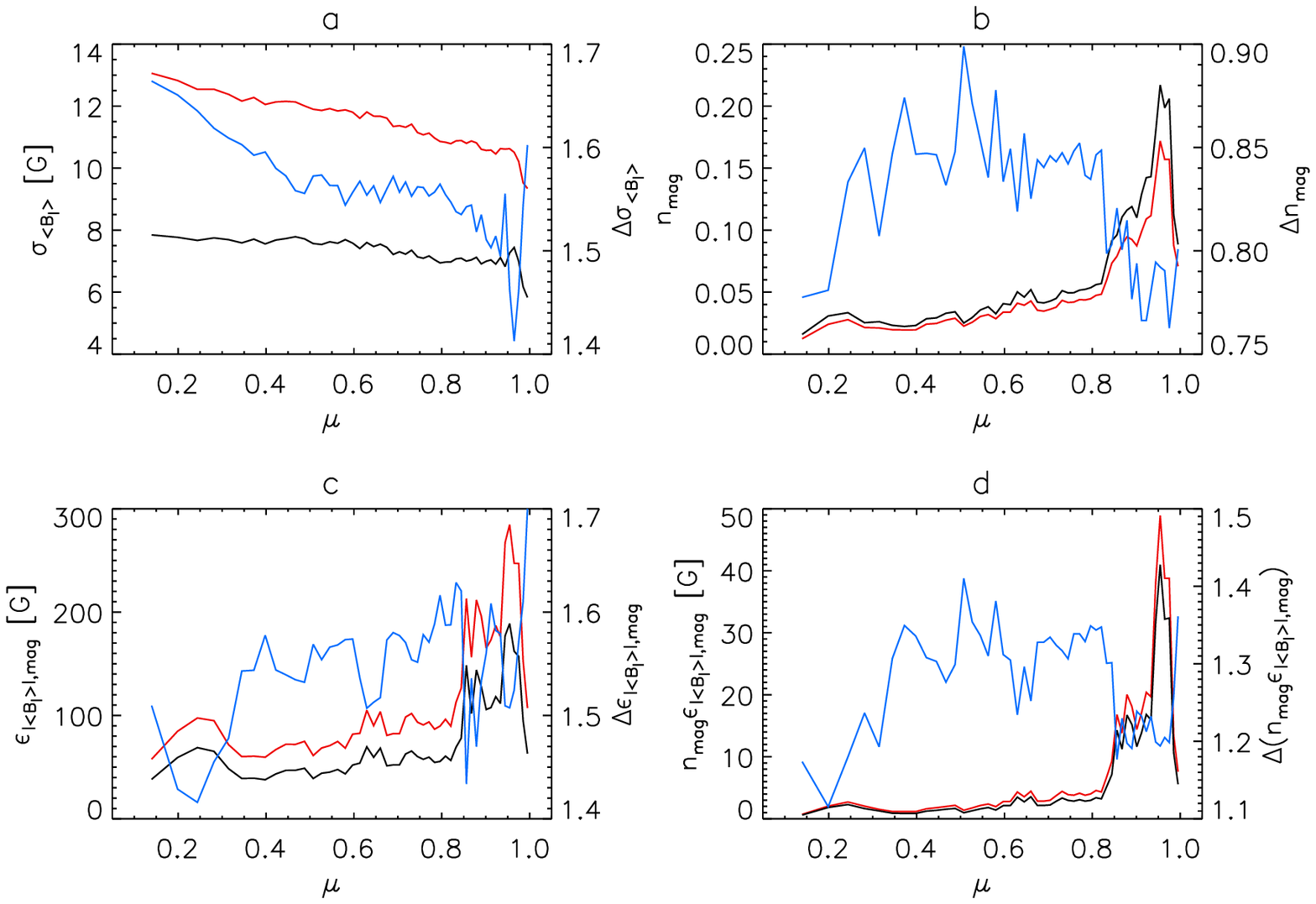}
\caption{Plotted as a function of $\mu$, a) the noise level of the 720-s longitudinal magnetogram, $\noise$, b) the proportion of image pixels counted as containing significant magnetic activity, $\nmag$, c) the mean $\bl$ of these points, $\mblmag$, d) and the product of $\nmag$ and $\mblmag$. Left axes: the uncorrected (black series) and restored with $\psftwo$ levels (red series). Right axes: the quotient of the restored and uncorrected values (blue series).}
\label{delta3}
\end{figure*}

We segmented the solar disc in the 720-s longitudinal magnetogram by $\mu$ (excluding points where $\mu<0.1$, $\sim1\%$ of the solar disc by area) into 50 annuli of equal area. Within each annulus, we computed the quantities listed below, plotted in Fig. \ref{delta3}. For each quantity, we derived the level in the uncorrected and restored magnetogram (black and red series, left axes), and the ratio of the restored and the uncorrected values (blue series, right axes), denoted by the $\Delta$ preffix.
\begin{itemize}
	\item The noise level, $\noise$ (Fig. \ref{delta3}a), given by the standard deviation of $\vbl$. The standard deviation was computed iteratively, points more than three standard deviations from the mean were excluded from succeeding iterations till convergence, to exclude magnetic activity. The variation of the noise level of HMI longitudinal magnetograms with position on the solar disc is dominated by a centre-to-limb increase \citep{liu12,yeo13}. It is therefore reasonable to represent the noise level within a given annulus by a single value of $\noise$. Image restoration increased the noise level, on average, by a factor of 1.6.
	\item The proportion of image pixels counted as containing significant magnetic activity, $\nmag$ (Fig. \ref{delta3}b), taken here as the points where $\bl>3\noise$.
	\item The mean $\bl$ of the image pixels counted as magnetic, $\mblmag$ (Fig. \ref{delta3}c).
	\item The product of $\nmag$ and $\mblmag$ (Fig. \ref{delta3}d). The quantity $\dflux$ represents the factor by which the apparent amount of line-of-sight magnetic flux changed from stray light removal.
\end{itemize}
Image restoration resulted in less image pixels being counted as magnetic (around $-10\%$ to $-25\%$, Fig. \ref{delta3}b), though the enhancement to the magnetogram signal (30$\%$ to 60$\%$, Fig. \ref{delta3}c) meant that there is an overall increase in the apparent amount of line-of-sight magnetic flux (10$\%$ to 40$\%$, Fig. \ref{delta3}d). Computing $\dflux$ taking all the annulus as a whole, the total amount of line-of-sight magnetic flux over the solar disc increased by a factor of 1.2.

\begin{figure}
\resizebox{\hsize}{!}{\includegraphics{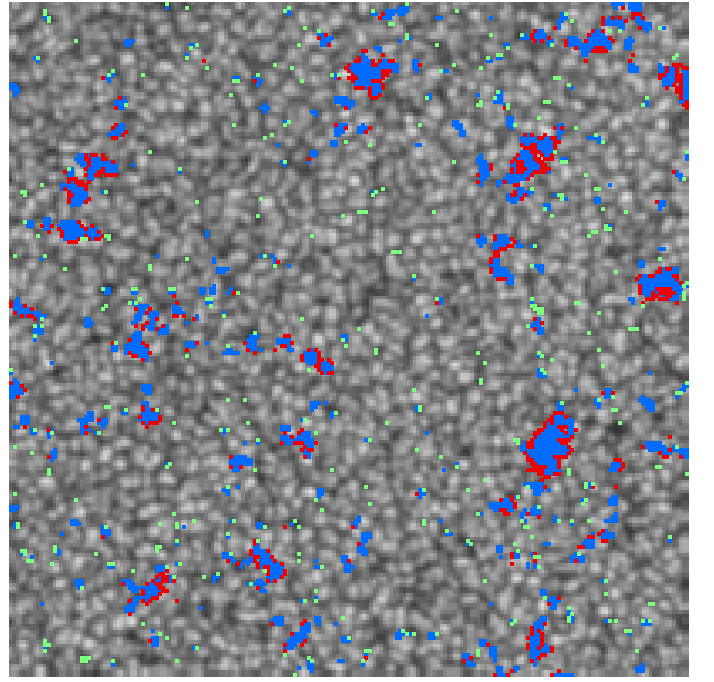}}
\caption{$201\times201$ pixel inset, centred on disc centre, of the 720-s continuum intensity image, with the points displaying $\bl>3\noise$ in the 720-s longitudinal magnetogram highlighted. Blue corresponds to the points realizing this condition in both the uncorrected and restored magnetogram, red the points fulfilling it only in the uncorrected data and green the points satisfying it only in the restored data. The grey scale is saturated at 0.8 and 1.2 times the mean quiet-Sun level.}
\label{delta4}
\end{figure}

In Fig. \ref{delta4}, we mark the location of image pixels counted as magnetic in the original and restored data, in a $201\times201$ pixel inset centred on the disc centre. Image restoration renders magnetic features spatially smaller as polarized radiation originating from these features, lost to the surrounding quiet Sun by aperture diffraction and stray light, is recovered (illustrated by the blue and red clusters). This change in the size of magnetic features likely depends on factors such as the surface area, and circumference to surface area ratio. The enhanced noise level also contributes to the smaller count in the restored data. Image restoration does recover some magnetic features smeared below the magnetogram signal threshold ($\bl=3\noise$) in the original data by instrumental scattered light (green clusters). Overall, less image pixels are counted as magnetic.

The overall increase in $\noise$ towards the limb is partly related to the increase in low-level magnetogram signal fluctuations from the ubiquitous weak horizontal magnetic fields in the quiet Sun internetwork \citep{lites96,lites08,beck09}, which obtain a line-of-sight component near the limb, and magnetic features foreshortening towards the background noise regime when approaching the limb. A probable cause of the overall centre-to-limb increase in $\Delta\noise$ is the enhancement of these true signal contributions to apparent noise.

In Fig. \ref{speg}a we display the change in magnetogram signal as a function of the original signal, in the quiet Sun field illustrated in Fig. \ref{delta4} and in a $201\times201$ pixel crop of an active region near the disc centre ($\mu>0.92$), represented by the green and black curves, respectively. This was computed as done in Sect. \ref{ssmcss3} for the sunspots and pores in Fig. \ref{spotinset}, the results of which are also plotted for comparison (blue and red curves). The only difference is here we binned the image pixels by the uncorrected $\bl$ in intervals of $40\:{\rm G}$ instead of $100\:{\rm G}$. As done in Sect. \ref{ssmcss1}, we minimised the influence of pores present in the active region field by excluding the image pixels where $\inorm<\iqp$ and points up to three pixels away from them. (There are no image pixels where $\inorm<\iqp$ in the quiet Sun field.) The result from the quiet Sun and active region fields represent the effect of image restoration on the magnetogram signal of quiet Sun network, and active region faculae, respectively.

As noted for sunspots and pores (Sect. \ref{ssmcss2}), the enhancement of the magnetogram signal of network and faculae from image restoration is highly correlated to the original level. This is possibly from the restoration enhancing the signal in the core of magnetic features, where it is also typically stronger, while depressing the signal in the fringes, from the recovery of polarized radiation scattered from the core to the fringes and surrounding quiet Sun, as discussed above and visible for the magnetic features depicted in Fig. \ref{contrastnfpeak}.

The effect of image restoration on network and faculae is also more pronounced than in sunspots and pores, in particular for network. This is likely related to the smaller spatial scale of these features, which makes them more susceptible to stray light, and consequently they respond more acutely to restoration, than sunspots and pores. The stronger response of network compared to faculae is probably due to the fact that they appear in smaller clusters and the restoration of small-scale mixed polarities in the quiet Sun smeared out by instrumental scattered light.

As image restoration affects different magnetic features differently, the overall effect on the apparent amount of magnetic flux fluctuates with prevailing magnetic activity. This is the likely reason neither $\Delta\nmag$, $\Delta\mblmag$ nor $\dflux$ exhibit any obvious trend with $\mu$, modulated by the magnetic features present within each annulus. Importantly, the relatively acute effect of image restoration on network, and the fact that the solar disc is, by area, predominantly quiet Sun means most of the increase in the measured amount of magnetic flux comes from the enhancement of these features. This is consistent with the findings of \cite{krivova04}.

\subsubsection{Line-of-sight velocity}
\label{ssmcss4}

Here we are interested in the part of measured line-of-sight velocities in HMI Dopplergrams arising from convective motions on the solar surface. To this end, we corrected the 720-s Dopplergram for the contribution by the rotation of the Sun and the relative velocity of SDO to the Sun\footnote{Given by the radial, heliographic west and north velocity of the spacecraft relative to the Sun listed in the data header.}, $\vobs$, following the procedure of \cite{welsch13}. Oscillatory motions associated with $p-$modes are largely undetectable in the 720-s Dopplergram data product as it is a weighted combination of filtergram observations from a 1350-s period, much longer than the period of these oscillations ($\sim5$ minutes).

The rotation rate of the Sun varies with heliographic latitude, $\Phi$. Let $\vrot$ denote the velocity of the surface of the Sun from its rotation. We first determined the Stonyhurst latitude and longitude at each image pixel within the solar disc. Then, we derived the $\vrot$ at each latitude from the differential rotation profile by \cite{snodgrass83},
\begin{equation}
w\left(\Phi\right)=2.902-0.464\sin^{2}{\Phi}-0.328\sin^{4}{\Phi},
\label{snodgrass}
\end{equation}
which relates angular velocity, $w=\frac{|{\vrot}|}{R_{\sun}\cos{\Phi}}$ ($R_{\sun}$ being the radius of the Sun in metres) to $\Phi$. The contribution by $\vrot$ and $\vobs$ to measured velocity at a given image pixel is then given by the projection of $\vrot$ and $\vobs$ onto the line-of-sight to the point, taking the small-angle approximation. The projection of $\vobs$, in this instance, varied between 500 and $800\:\mps$ with position on the solar disc. The significant magnitude and variation of this with disc position arises from SDO's geosynchronous orbit about Earth and Earth's orbit about the Sun.

Let $\vl$ represent the signed Dopplergram signal, the mean line-of-sight component of the vector velocity over a given image pixel.

\begin{figure}
\resizebox{\hsize}{!}{\includegraphics{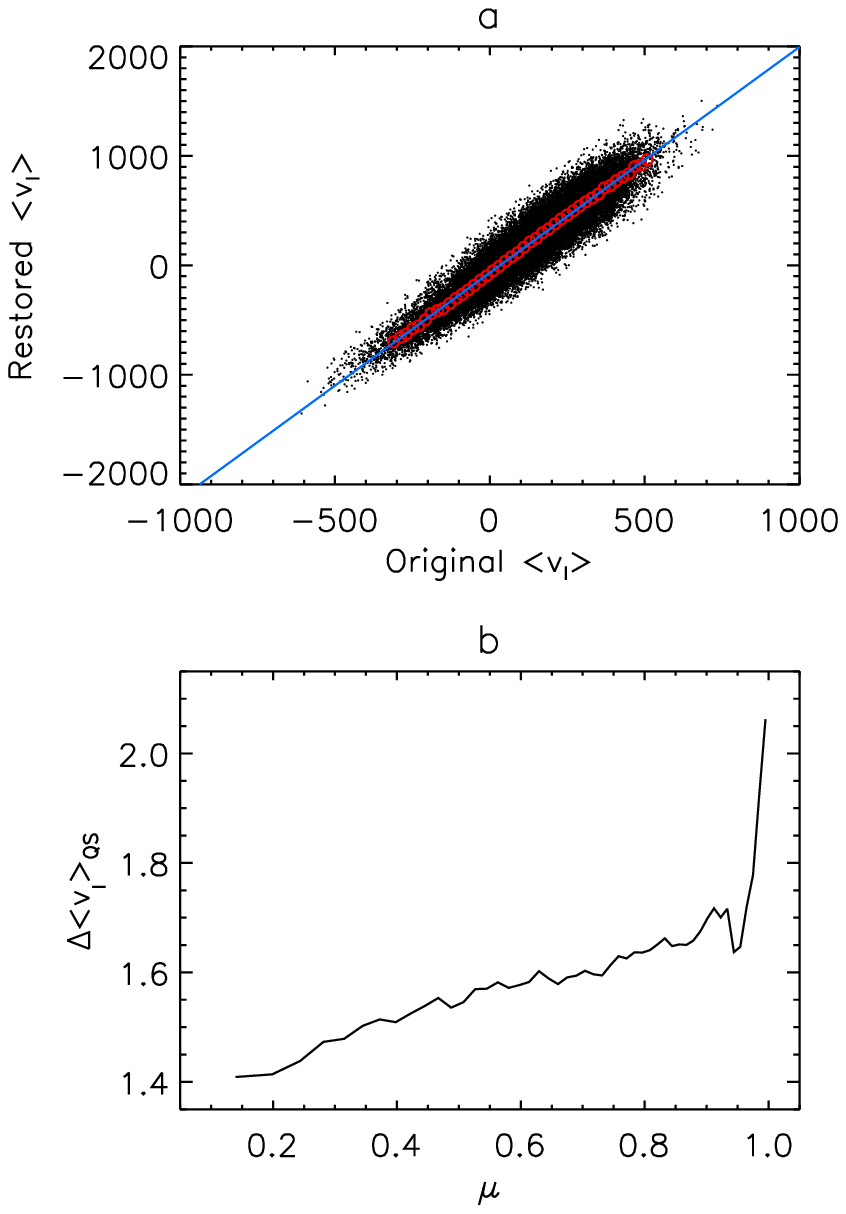}}
\caption{a) Scatter plot of the Dopplergram signal, $\vl$ in the quiet Sun, near disc centre ($\mu>0.99$), in the restored (with $\psftwo$) and uncorrected versions of the 720-s Dopplergram. The red circles denote the bin-median (from binning the points by the uncorrected $\vl$ in intervals of $20\:\mps$) and the blue line the corresponding linear regression fit. b) The factor enhancement of $\vl$ in the quiet Sun from image restoration, $\dvl$ as a function of $\mu$.}
\label{deltav}
\end{figure}

As done in Sect. \ref{ssmcss3}, we segmented the solar disc by $\mu$, into 50 equal annuli, excluding points where $\mu<0.1$. The image pixels where $\bmu>10{\rm G}$ in the 720-s longitudinal magnetogram were masked, leaving quiet Sun. Within each annulus, we binned the unmasked points by the uncorrected $\vl$ in intervals of $20\:\mps$, and retrieved the median original and restored $\vl$ within each bin, ignoring bins with less than 100 points. The factor enhancement of $\vl$ from image restoration, $\dvl$ is then given by the slope of the linear regression fit to the restored bin-median $\vl$ against the original, illustrated for the disc centremost interval of $\mu$ ($\mu>0.99$) in Fig. \ref{deltav}a. By performing this computation over small intervals of $\mu$, we avoid introducing scatter or bias in $\dvl$ from the $\mu$ dependence of the convective blueshift of the spectral line.

The enhancement of the Dopplergram signal in the quiet Sun from image restoration with $\psftwo$ is significant and exhibits an acute CLV; $\dvl$ increases monotonically with $\mu$, from about 1.4 near the limb to 2.1 at disc centre (Fig. \ref{deltav}b).

In looking only at the quiet Sun, we excluded phenomena localised in active regions (for example, Evershed flow in sunspots), leaving signal largely from granulation and supergranulation. Supergranulation flows are largely horizontal. The line-of-sight velocities of supergranulation flows are thus greatest near the limb and diminishes towards disc centre from foreshortening. In contrast, the apparent line-of-sight velocities of granulation flows diminish towards the limb. Approaching the limb, granulation is more and more difficult to resolve from foreshortening and the line-of-sight increasingly crossing into multiple granulation cells. The typical diameter of granulation and supergranulation cells is about $1\:{\rm Mm}$ and $30\:{\rm Mm}$, respectively. Granulation is therefore more affected by stray light and experiences greater signal enhancement from image restoration than supergranulation. The observed CLV of $\dvl$ is consistent with the converse CLV of the line-of-sight velocities of granulation and supergranulation flows, and the stronger effect of image restoration on granulation compared to supergranulation.

By correcting both the original and restored Dopplergram for differential rotation with Eqn. \ref{snodgrass}, we had implicitly assumed that this component of measured line-of-sight velocity is not significantly changed by stray light or its removal. Given the line-of-sight component of $\vrot$ varies gradually across the solar disc, this is true except very close to the limb. Therefore, the effect on this analysis is likely minute and confined to the annuli closest to the limb.

The pronounced Dopplergram signal enhancement effected by image restoration with $\psftwo$ could have an impact on the characterization of plasma flows in the solar surface with HMI data. In this study we will not attempt to examine the effects of image restoration on apparent Doppler shifts in active regions, $p-$mode oscillations (detectable in the 45-s Dopplergram data product of the front CCD) or the individual physical processes driving plasma motion on the solar surface.

\section{Variation of the PSF within the FOV, between the HMI CCDs and with time}
\label{psfdep}

Derived from observations of Venus in transit recorded on the side CCD, $\psftwo$ characterizes the stray light in the employed data, at the position in the FOV occupied by the venusian disc. Here we discuss the applicability of $\psftwo$ to the entire FOV, to data from the front CCD and, importantly, from other times.

\subsection{Variation of the PSF with FOV position}
\label{psfdep1}

\begin{figure}
\resizebox{\hsize}{!}{\includegraphics{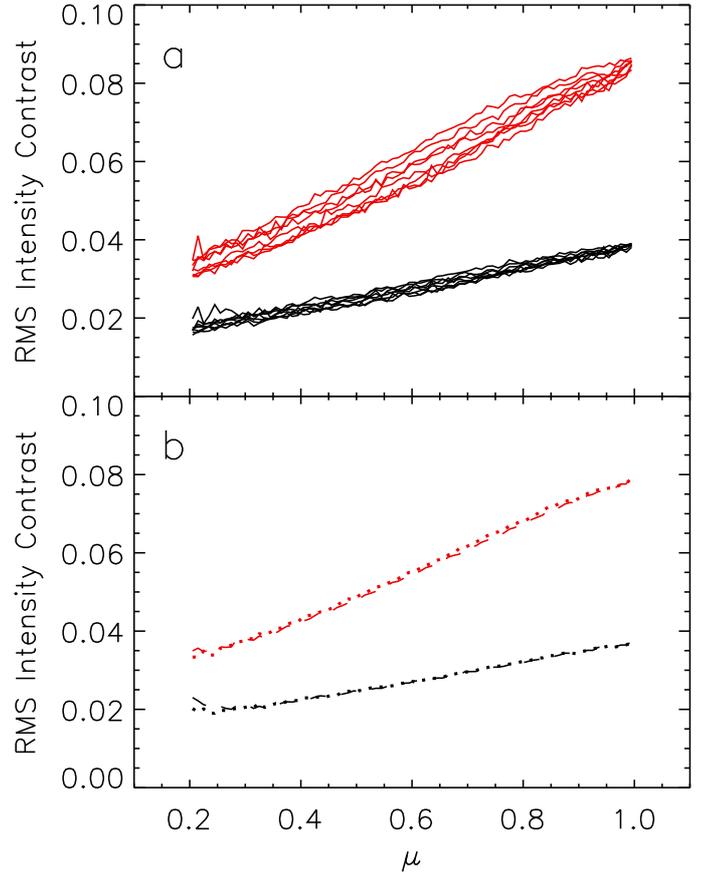}}
\caption{RMS intensity contrast of the quiet Sun as a function of $\mu$, before (black lines) and after (red) image restoration with $\psftwo$. a) In the test continuum filtergram, the solar disc segmented into eight equal sectors. b) In the near-simultaneous filtergrams of similar bandpass and polarization from the side (dotted lines) and front (dashed) CCD.}
\label{contrastqs2}
\end{figure}

Taking the test continuum filtergram (from Sect. \ref{gc}), we segmented the solar disc into eight equal sectors and computed the RMS intensity contrast of the quiet Sun within each sector. The level before and after image restoration with $\psftwo$ are illustrated in Fig. \ref{contrastqs2}a. There is scatter in the RMS contrast between the sectors which is more pronounced in the restored data. This enhanced divergence is, at least in part, caused by the fact that we restored the entire solar disc with a single PSF when the true PSF varies from sector to sector.

The scatter in RMS contrast is, however, at least to the extent tested, relatively small in comparison to the absolute level. More importantly, the divergence between the sectors in the restored data (red curves) is small compared to the difference between the restored and the original data (black curves). This implies that the inhomogeneity introduced by applying a single PSF to the entire FOV is small in comparison to the contrast enhancement. Nonetheless, for sensitive measurements, care should be taken, where possible, to average measurements from different positions in the FOV after deconvolution with the PSF deduced here.

Next, we looked at the variation, over the solar disc, of the effect of image restoration with $\psftwo$ on the 720-s longitudinal magnetogram from Sect. \ref{ssmcss}. Sampling the solar disc at 16-pixel intervals in both the north-south and east-west directions, we centred a $401\times401$ pixel window over each sampled point and took the mean $\bl$ of all the solar disc pixels within the window, denoted $\mbl$. Let $\dbl$ represent the ratio of $\mbl$ in the restored and uncorrected data, representing the factor enhancement to $\mbl$ from the restoration.

\begin{figure}
\resizebox{\hsize}{!}{\includegraphics{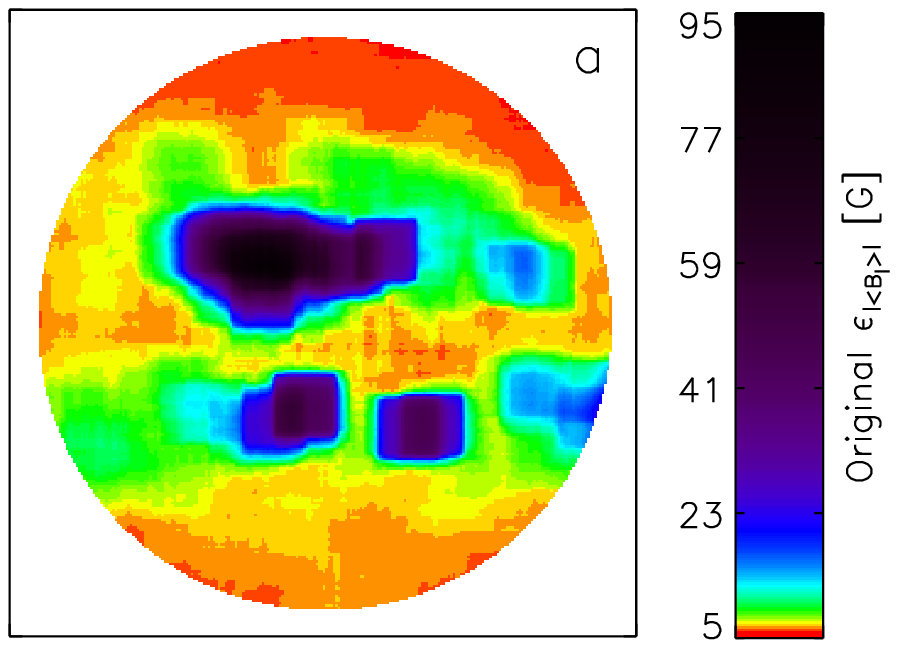}}
\resizebox{\hsize}{!}{\includegraphics{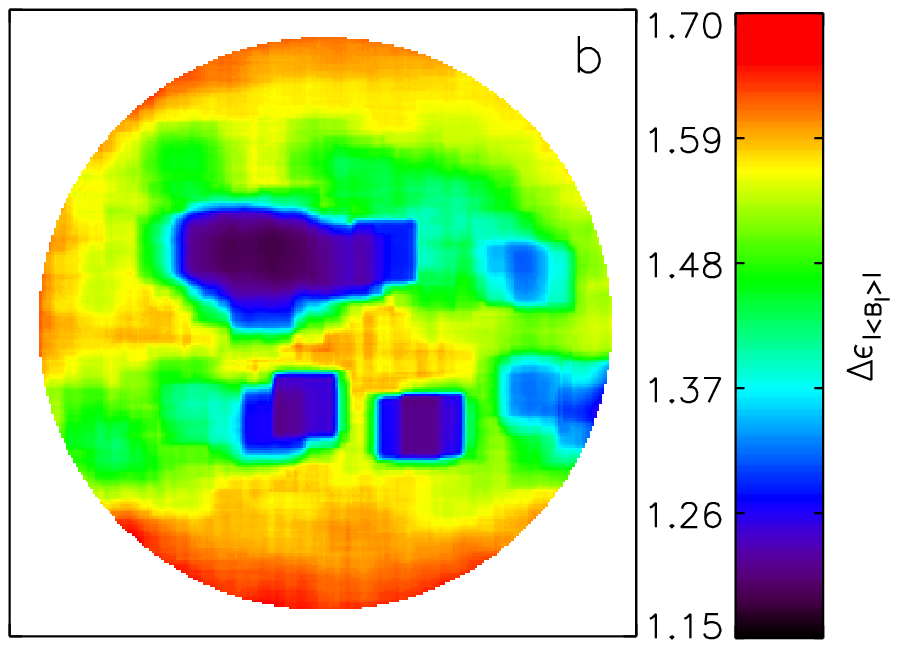}}
\resizebox{\hsize}{!}{\includegraphics{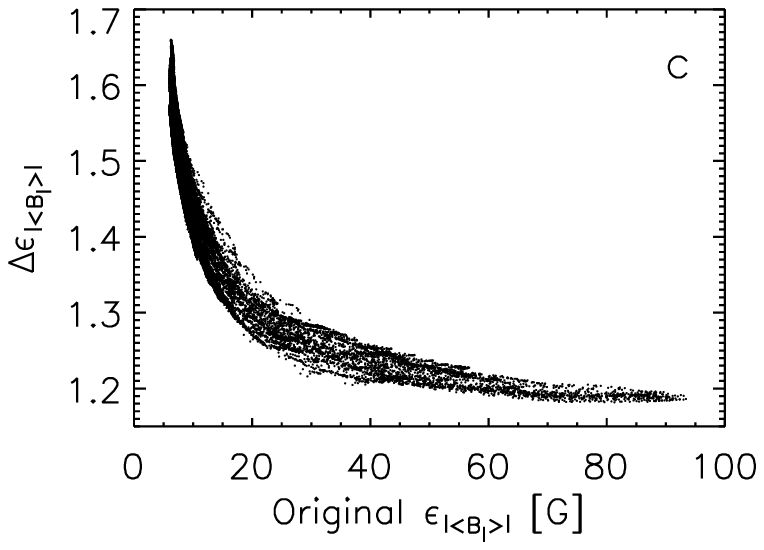}}
\caption{a) Mean unsigned magnetogram signal, $\mbl$ of the original (not corrected for stray light) 720-s longitudinal magnetogram. a) Factor enhancement of $\mbl$ from image restoration with $\psftwo$, $\dbl$. c) Scatter plot of $\dbl$ against the original $\mbl$.}
\label{deltamag}
\end{figure}

Expectedly, $\dbl$ varied with position on the solar disc (Fig. \ref{deltamag}b). This variation is driven by:
\begin{itemize}
	\item differences in the magnetic activity present in the sampling window (image restoration affects different magnetic features differently, Sect. \ref{ssmcss3}) and
	\item fluctuations in the effect of image restoration from the variation of the stray light behaviour of the instrument across the FOV.
\end{itemize}
The scatter plot of $\dbl$ versus the original $\mbl$ reveals an inverse relationship between the two quantities (Fig. \ref{deltamag}c). The cause of this correlation is that the restoration enhances the magnetogram signal in the quiet Sun more strongly than in active regions (Fig. \ref{speg}a). The relatively weak spread of the scatter plot suggests the variation in $\dbl$ with disc position is dominantly due to the inhomogeneous distribution of magnetic activity (Fig. \ref{deltamag}a) and the diverging effects of restoration on different magnetic features. This is further evidence that it is a reasonable approximation to restore the entire FOV of HMI for instrumental scattered light with a single PSF.

\subsection{Variation of the PSF between the HMI CCDs and with time}
\label{psfdep2}

In Fig. \ref{contrastqs2}b we show the RMS intensity contrast of the quiet Sun in two near-simultaneous filtergrams, one from each CCD, taken less than one minute apart. The two filtergrams were recorded about an hour after Venus left the solar disc (at around 05:36 UTC, June 6, 2012), shortly after the side CCD resumed collection of the regular filtergram sequence.

The two near-simultaneous filtergrams were taken in the same bandpass (-172 m\AA{} from line centre) and polarization (Stokes $I-V$). Any disparity in the RMS contrast between the two would arise mainly from differences in the performance of the two CCDs, including the stray light behaviour. The RMS contrast in the two filtergrams is very similar, even after image restoration with $\psftwo$ (red curves). Since we expect any disparity due to differences in the stray light behaviour of the two CCDs to be amplified by the application of the side CCD PSF to a front CCD filtergram, this agreement is encouraging.

Comparing the close similarity in the RMS intensity contrast of the quiet Sun in the two near-simultaneous filtergrams to the scatter between different FOV positions in the test continuum filtergram (top panel), it appears that the difference in the stray light behaviour of the two CCDs is much smaller than the variation with position in the FOV of the side CCD. The side and front CCDs are identical and share a common optical path \citep{schou12}, it is therefore within reason that the stray light behaviour, even the variation of the PSF with position in the FOV, is broadly similar.

The SDO satellite is in a geosynchronous orbit. Since the HMI commenced regular operation (May 1, 2010), the side and front CCDs recorded a continuum filtergram each during the daily pass through orbital noon and midnight. For the purpose of investigating the stability of the stray light behaviour of the instrument over time, we examined the solar aureole, the intensity observed outside the solar disc arising from instrumental scattered light, in these daily data\footnote{The solar aureole is not to be confused with the aureole about the venusian disc when it is in transit, discussed in Sect. \ref{venusatm}.}. We examined 2307 orbital noon and midnight continuum filtergrams from each CCD, spanning the 1157-day period of May 1, 2010 to June 30, 2013.

\begin{figure*}
\centering
\includegraphics[width=17cm]{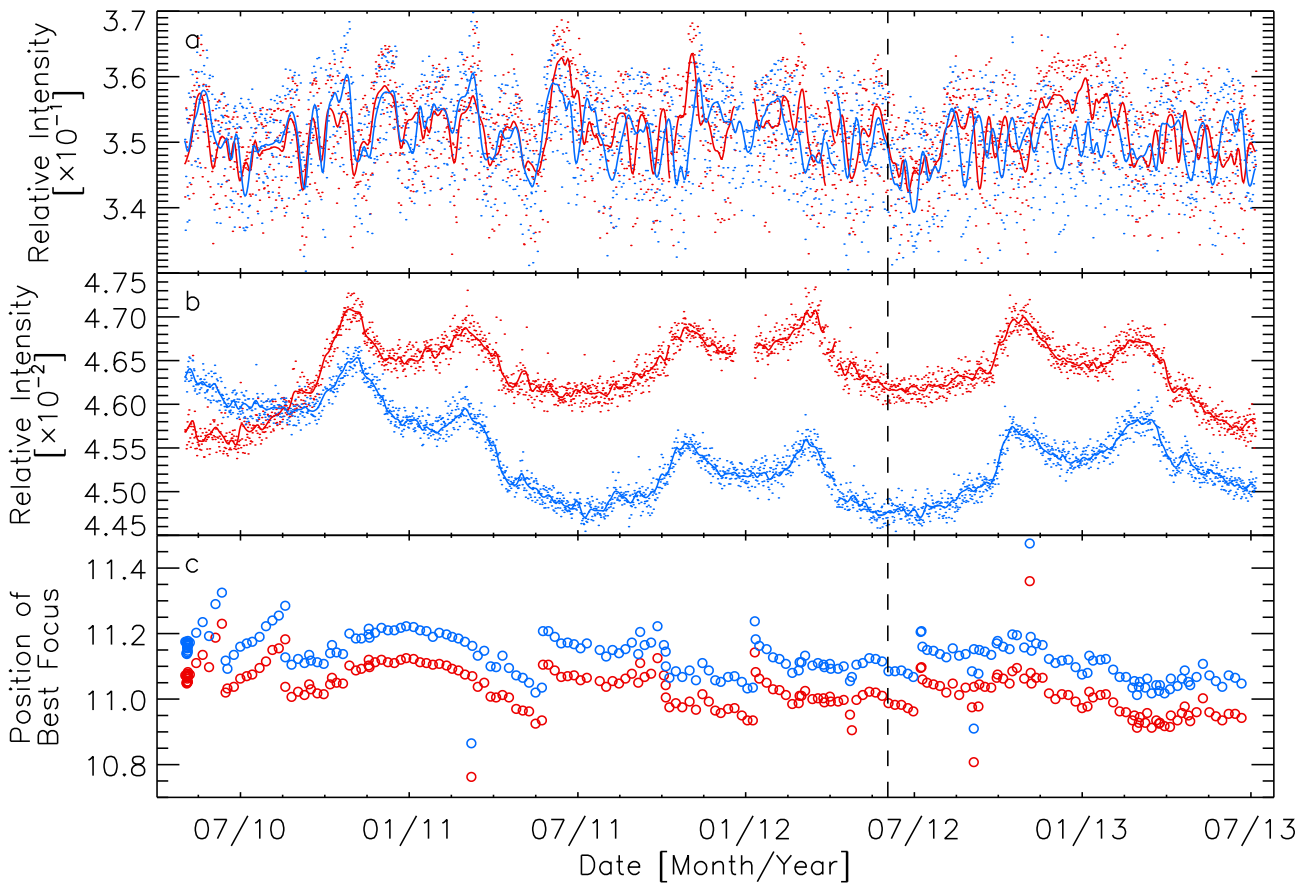}
\caption{Radial intensity of the solar aureole, at distances of a) 1 and b) 10 arcsec from the edge of the solar disc, in the orbital noon and midnight continuum filtergrams from the side (red) and front (blue) CCDs. The value from each filtergram is normalized to the level at the edge of the solar disc. The dots represent the measured values and the curves the smoothed time series. Segments of measured values spaced more than one day apart were treated as separate time series, giving the gaps in the curves. c) The position of best focus, in units of focus steps (see text), from the (approximately) weekly focus calibration of the side (red circles) and front CCDs (blue circles). The dashed line marks the time of the transit of Venus.}
\label{analysistime}
\end{figure*}

For each continuum filtergram, we averaged the solar aureole over all azimuths. From the resulting radial intensity profile, we determined the intensity of the solar aureole at distances of 1 and 10 arcsec from the edge of the solar disc. The derived intensities, normalized to the limb level, are expressed in Figs. \ref{analysistime}a and \ref{analysistime}b, respectively.

We excluded the points that are spurious or from continuum filtergrams with missing pixel values, leaving 2248 and 2288 points from the side and front CCDs respectively. To show up the broad trend with time we interpolated each time series at one day intervals and smoothed the result by means of binomial smoothing \citep{marchand83}. The time variation in the relative (to the limb level) intensity of the solar aureole at 1 and 10 arcsec from the limb reflects changes to the shape of the PSF near the core and in the wings respectively.

There is clear point-to-point fluctuation in the relative intensity of the solar aureole at 1 arcsec from the limb. This is due to the fact that near the limb, the intensity of the solar aureole decays rapidly with radial distance such that small variations in the width of the core of the PSF show up as large swings in the relative intensity at a fixed distance from and near the limb. Overall, the magnitude of the scatter, given by the standard deviation, is below $2\%$ of the mean level. Also, the side and front CCD series are relatively stable and similar, with no overt long term trends or divergence from one another. This suggests that the width of the core of the PSFs of the two CCDs are, to the extent tested, broadly similar and constant over the period examined.

For the relative intensity of the solar aureole at 10 arcsec from the limb, there is an approximately synchronous periodic variation between the two CCDs. There is also a gradual overall drift between May 2010 and July 2011, and between August 2012 and June 2013. The fluctuation of the two time series and the divergence between them is minute, less than $5\%$ of the overall level. The time variation of, and difference between the wings of the PSFs of the two CCDs, implied by these fluctuations, has likely little effect on image contrast. Even with the $\sim3\%$ offset between the front and side CCD time series around the time of the Venus transit, the RMS intensity contrast of the quiet Sun in the near-simultaneous filtergrams from the two CCDs (taken about an hour after the end of the transit) is practically identical (Fig. \ref{contrastqs2}b).

Going from 1 to 10 arcsec from the limb, the time dependence of the relative intensity of the solar aureole changes gradually from the trend seen at 1 arcsec to that seen at 10 arcsec. Beyond 10 arcsec the variation with time does not change significantly. The relative intensity of the solar aureole at 1 and 10 arcsec from the limb therefore constitute a reasonable representation of variation in the shape of the PSFs of the two CCDs with time.

From the near identity of the RMS intensity contrast of the quiet Sun in the near-simultaneous filtergrams from the two CCDs (Fig. \ref{contrastqs2}b) and the minute time variation of the PSFs of the two CCDs implied by the relative stability of the shape of the solar aureole in daily data (Figs. \ref{analysistime}a and \ref{analysistime}b), we assert that $\psftwo$ can be applied to observations from both CCDs for the period examined (May 1, 2010 to June 30, 2013) without introducing significant error.

This assertion is consistent with the state of the focus of the instrument over the period examined. The PSF of HMI, like any optical system, is strongly modulated by the focus. The instrument is maintained in focus by varying the heating of the front window and the position of two five-element optical wheels \citep{schou12}. By varying the elements of the two optical wheels placed in the optical path, they allow the focus to be adjusted in 16 uniform, discrete steps, each corresponding to about two-thirds of a depth of focus. In Fig. \ref{analysistime}c we plot the position of best focus, in units of the focus steps, from the regular (roughly weekly) focus calibration of the sensor, over the same period as the daily data. The focus of the two CCDs are remarkably similar and stable, differing from one another and varying over the period of interest by much less than a depth of focus. The contribution by focus to the variation of the PSF between the two CCDs and with time is most probably minute.

\section{Summary}
\label{summary}

In this paper we present an estimate of the PSF of the SDO/HMI instrument. The PSF was derived from observations of Venus in transit. We convolved a simple model of the venusian disc and solar background with a guess PSF iteratively, optimizing the agreement between the result of the convolution and observation. We modelled the PSF as the linear sum of five Gaussian functions, the amplitude of which we allowed to vary sinusoidally with azimuth. This azimuthal variation was necessary to reproduce the observations accurately. Recovering the anisotropy of the PSF was also shown to be important for the proper removal of stray light from HMI data by the deconvolution with the PSF. The interaction between solar radiation and the venusian atmosphere is complex and not straightforward to account for. The result is a conservative estimate of the PSF, similar in width to the ideal diffraction-limited PSF in the core but with more extended wings.

The PSF was derived with data from one of the two identical CCDs in the sensor. It therefore represents the stray light behaviour of that particular CCD, at the time of the transit of Venus, at the position in the FOV occupied by the venusian disc in the employed observations.

Comparing the apparent granulation contrast in different parts of a single image, we showed that although the variation in the stray light behaviour of the instrument with position in the FOV introduces uncertainty to measured contrast, amplified by restoring observations with a single PSF, the scatter is relatively minute and will likely have little quantitative influence if care is taken to average measurements from multiple FOV positions. This was confirmed by an examination of the uniformity, over the FOV, of the effect of image restoration on the 720-s longitudinal magnetogram data product.

The time variation of the shape of the solar aureole in daily data was taken as an indication of PSF changes over the period examined (May 1, 2010 to June 30, 2013). Based on the relatively weak time variation of the aureole, and the similarity of the aureole and granulation contrast in data from the two CCDs, we assert that the PSF derived here can be applied to observations from both CCDs over the period examined without introducing significant error.

Apparent granulation contrast, given here by the RMS intensity contrast of the quiet Sun, in HMI continuum observations restored by the deconvolution with the PSF, exhibit reasonable agreement with that in artificial images generated from a 3D MHD simulation at equal spatial sampling. This demonstrates that the PSF, though an approximation, returns a competent estimate of the aperture diffraction and stray light-free contrast. The restoration enhanced the RMS intensity contrast of the quiet Sun by a factor of about 1.9 near the limb ($\mu=0.2$), rising up to 2.2 at disc centre.

We also illustrated the effect of image restoration with the PSF on the 720-s Dopplergram and longitudinal magnetogram data products, and the apparent intensity of magnetic features.
\begin{itemize}
	\item For small-scale magnetic concentrations, image restoration enhanced the intensity contrast in the continuum and core of the Fe I 6173 \AA{} line by a factor of about 1.3, and the magnetogram signal by a factor of about 1.7.
	\item Magnetic features in the longitudinal magnetogram are rendered smaller, as polarized radiation smeared onto surrounding quiet Sun by instrumental scattered light is recovered.
	\item Image restoration increased the apparent amount of magnetic flux above the noise floor by a factor of about 1.2, mainly in the quiet Sun. This may be, in part, from the recovery of magnetic flux in opposite magnetic polarities lying close to one another partially cancelled out by stray light.
	\item The influence of image restoration on sunspots and pores varied strongly, as expected, within a given feature and between features of different sizes.
	\item Line-of-sight velocity due to plasma motions on the solar atmosphere increases by a factor of about 1.4 to 2.1. The variation comes from the restoration enhancing granulation flows more strongly than larger scale supergranulation flows. Given the spatial scale dependence, the effect on Doppler shifts from large scale motions such as the meridional flow and differential rotation is probably minute.
\end{itemize}
The pronounced effect of image restoration on the apparent radiant, magnetic and motional properties of solar surface phenomena could have a significant impact on the interpretation of HMI observations. For instance, the observation that restoring HMI longitudinal magnetograms renders magnetic features smaller while boosting the magnetogram signal, and the increase in the amount of magnetic flux is dominantly in the quiet Sun, can influence models of variation in solar irradiance based on HMI data. Specifically, with models that relate the contribution by small-scale magnetic concentrations to solar irradiance variations to the number density and magnetogram signal \citep[for example,][]{wenzler06,foukal11,ball12}.

\begin{acknowledgements}
The authors would like to express their gratitude to the SDO/HMI team, for providing the data set, and all the useful discussions and support rendered over the course of this study, with special mention to Philip Scherrer and Jesper Schou. The authors would also like to thank the members of the Solar Lower Atmosphere and Magnetism group at the Max-Planck Institut f\"{u}r Sonnensystemforschung, in particular Achim Gandorfer, for their interest in this work, and all the helpful advice and comments. K.L.Y. acknowledges postgraduate fellowship of the International Max Planck Research School on Physical Processes in the Solar System and Beyond. This work has been partly supported by WCU grant No. R31-10016 funded by the Korean Ministry of Education, Science and Technology.
\end{acknowledgements}

\end{document}